\documentclass[preprint,12pt]{elsarticle}

\usepackage{graphicx}
\usepackage{amssymb}
\usepackage{longtable}
\usepackage{multirow}
\usepackage{appendix}
\usepackage[T1]{fontenc}
\usepackage{lscape}

\usepackage{multicol}
\usepackage{graphicx}
\usepackage{lineno,hyperref}
\usepackage{xcolor}
\usepackage{enumerate}
\usepackage{enumitem} 
\usepackage{subcaption}
\usepackage{float}
\usepackage{breakurl}
\usepackage{dcolumn}
\usepackage{latexsym}
\usepackage{placeins}
\usepackage[utf8]{inputenc}
\usepackage{amsmath}
\usepackage[boxruled,vlined,linesnumbered]{algorithm2e}
\usepackage{mdframed}
\mdfdefinestyle{style1}{leftmargin=0.5cm,rightmargin=0.5cm,skipbelow=0.5cm,skipabove=0.2cm,
   innertopmargin=2pt}

\usepackage{hyperref}
\hypersetup{
    colorlinks=true,
    linkcolor=blue,
    filecolor=magenta,      
    urlcolor=cyan,
}
\usepackage[a4paper, total={6in, 10in}]{geometry}

\journal{Journal Name}

\begin{document}

\begin{frontmatter}

\title{An Evaluation of Monte Carlo-Based Hyper-Heuristic for Interaction Testing of Industrial Embedded Software Applications}

\author{Bestoun S. Ahmed}
\address{Department of Mathematics and Computer Science, Karlstad University\\ 651 88 Karlstad, Sweden\\ email: bestoun@kau.se}
\author{Eduard Enoiu}
\address{M\"alardalen University, V\"aster{\aa}s, Sweden\\ email: {\{firstname.lastname@mdh.se\}}}
\author{Wasif Afzal}
\address{M\"alardalen University, V\"aster{\aa}s, Sweden\\ email: {\{firstname.lastname@mdh.se\}}}
\author{Kamal Z. Zamli}
\address{Faculty of Computing \\ University Malaysia Pahang, Pekan, Malaysia.\\ email: {kamalz@ump.edu.my}}

\begin{abstract}
Hyper-heuristic is a new methodology for the adaptive hybridization of meta-heuristic algorithms to derive a general algorithm for solving optimization problems. This work focuses on the selection type of hyper-heuristic, called the Exponential Monte Carlo with Counter (EMCQ). Current implementations rely on the memory-less selection that can be counterproductive as the selected search operator may not (historically) be the best performing operator for the current search instance. Addressing this issue, we propose to integrate the memory into EMCQ for combinatorial $t-wise$ test suite generation using reinforcement learning based on the Q-learning mechanism, called Q-EMCQ. The limited application of combinatorial test generation on industrial programs can impact the use of such techniques as Q-EMCQ. Thus, there is a need to evaluate this kind of approach against relevant industrial software, with a purpose to show the degree of interaction required to cover the code as well as finding faults. We applied Q-EMCQ on 37 real-world industrial programs written in Function Block Diagram (FBD) language, which is used for developing a train control management system at Bombardier Transportation Sweden AB. The results of this study show that Q-EMCQ is an efficient technique for test case generation. Additionally, unlike the t-wise test suite generation, which deals with the minimization problem, we have also subjected Q-EMCQ to a maximization problem involving the general module clustering to demonstrate the effectiveness of our approach. The results show the Q-EMCQ is also capable of outperforming the original EMCQ as well as several recent meta/hyper-heuristic including Modified Choice Function (MCF), Tabu High Level Hyper-Heuristic(HHH), Teaching Learning based Optimization (TLBO), Sine Cosine Algorithm (SCA) and Symbiotic Optimization Search (SOS) in clustering quality within comparable execution time.

\end{abstract}


\end{frontmatter}

\section{Introduction}

Despite their considerable success, meta-heuristic algorithms have been adapted to solve specific problems based on some domain knowledge. Some examples of recent meta-heuristic algorithms include Sooty Tern Optimization Algorithm (STOA) \cite{REF-A}, Farmland Fertility Algorithm (FF) \cite{REF-B}, Owl Search Algorithm (OSA) \cite{REF-C}, Human Mental Search (HMS) \cite{REF-D} and Find-Fix-Finish-Exploit-Analyze (F3EA) \cite{REF-E}. Often, these algorithms require significant expertise to implement and tune; hence, their standard versions are not sufficiently generic to adapt to changing search spaces, even for the different instances of the same problem. Apart from this need to adapt, existing research on meta-heuristic algorithms has also not sufficiently explored the adoption of more than one meta-heuristic to perform the search (termed \textit{hybridization}). Specifically, the exploration and exploitation of existing algorithms are limited to using the (local and global) search operators derived from a single meta-heuristic algorithm as a basis. In this case, choosing a proper combination of search operators can be the key to achieving good performance as hybridization can capitalize on the strengths and address the deficiencies of each algorithm collectively and synergistically.

Hyper-heuristics have recently received considerable attention for addressing some of the above issues~\cite{Tsai2014,SABAR2015225}. Specifically, hyper-heuristic represents an approach of using (meta)-heuristics to choose (meta)-heuristics to solve the optimization problem at hand~\cite{Burke2003}. Unlike traditional meta-heuristics, which directly operate on the solution space, hyper-heuristics offer flexible integration and adaptive manipulation of complete (low level) meta-heuristics or merely the partial adoption of a particular meta-heuristic search operator through non-domain feedback. In this manner, hyper-heuristic can evolve its heuristic selection and acceptance mechanism in searching for a good-quality solution.

This work is focusing on a specific type of hyper-heuristic algorithm, called the Exponential Monte Carlo with Counter (EMCQ)~\cite{SABAR2015225,Batat2014}. EMCQ adopts a simulated annealing like~\cite{Kirkpatrick671} reward and punishment mechanism to adaptively choose the search operator dynamically during run-time from a set of available operators. To be specific, EMCQ rewards a good performing search operator by allowing its re-selection in the next iteration. Based on decreasing probability, EMCQ also rewards (and penalizes) a poor performing search operator to escape from local optima. In the current implementation, when a poor search operator is penalized, it is put in the tabu list, and EMCQ will choose a new search operator from the available search operators randomly. Such memory-less selection can be counter-productive as the selected search operator may not (historically) be the best performing operator for the current search instance. For this reason, we propose to integrate the memory into EMCQ using reinforcement learning based on the Q-learning mechanism, called Q-EMCQ.

We have adopted Q-EMCQ for combinatorial interaction $t-wise$ test generation (where $t$ indicates the interaction strength). While there is already significant work on adopting hyper-heuristic as a suitable method for $t-wise$ test suite generation (see, e.g.,~\cite{Zamli:2016:TSH,Zamli:2017:ESH})), the main focus has been on the generation of minimal test suites. It is worthy of mentioning here that, in this work, our main focus is not to introduce new bounds for the $t-wise$ generated test suites. Rather we dedicate our efforts on assessing the effectiveness and efficiency of the generated $t-wise$ test suites against real-world programs being used in industrial practice. Our goal is to push towards the industrial adoption of $t-wise$ testing, which is lacking in numerous studies on the subject. We, nevertheless, do compare the performance of Q-EMCQ against the well-known benchmarks using several strategies, to establish the viability of Q-EMCQ for further empirical evaluation using industrial programs. In the empirical evaluation part of this paper, we rigorously evaluate the effectiveness and efficiency of Q-EMCQ for different degrees of interaction strength using real-world industrial control software used for developing the train control management system at Bombardier Transportation Sweden AB. To demonstrate the generality of Q-EMCQ, we have also subjected Q-EMCQ a maximization problem involving the general module clustering. Q-EMCQ entails the best overall performance on the clustering quality and within a comparable execution time when compared to competing hyper-heuristics (MCF, and Tabu HHH) and meta-heuristics (EMCQ, TLBO, SCA, and SOS). Summing up, this paper makes the following contributions:
 
This paper makes the following contributions:

\begin{enumerate}
\item A novel Q-EMCQ hyper-heuristic technique that embeds the Q-learning mechanism into EMCQ, providing a memory of the performance of each search operator for selection. The implementation of Q-EMCQ establishes a unified strategy for the integration and hybridization of Monte Carlo-based exponential Metropolis probability function for meta-heuristic selection and acceptance mechanism with four low-level search operators consisting of cuckoo’s Levy flight perturbation operator~\cite{Suash2009}, flower algorithm’s local pollination, and global pollination operator~\cite{YangShi2012} as well as Jaya’s search operator~\cite{Rao2016}.

\item  An industrial case study, evaluating $t-wise$ test suite generation in terms of cost (i.e., using a comparison of the number of test cases) and effectiveness (i.e., using mutation analysis).

\item Performance assessment of Q-EMCQ with contemporary meta/hyper-heuristics for maximization problem involving general module clustering problem.

\end{enumerate}

\section{Theoretical Background and an Illustrative Example}

Covering array ($CA$) is a mathematical object to represent the actual set of test cases based on $t-wise$ coverage criteria (where $t$ represents the desired interaction strength). $CA (N; t, k, v)$, also expressed as $CA (N; t, v^k)$, is a combinatorial structure constructed as an array of $N$ rows and $k$ columns on $v$ values such that every $N \times t$ sub-array contains all ordered subsets from the $v$ values of size $t$ at least once. Mixed covering array $(MCA) (N; t, k, (v_1, v_2,\ldots \ v_k))$ or $MCA (N; t, k, v^k)$ may be adopted when the number of component values varies. 

\begin{figure*}

\centering

\includegraphics[width= 4 in]					{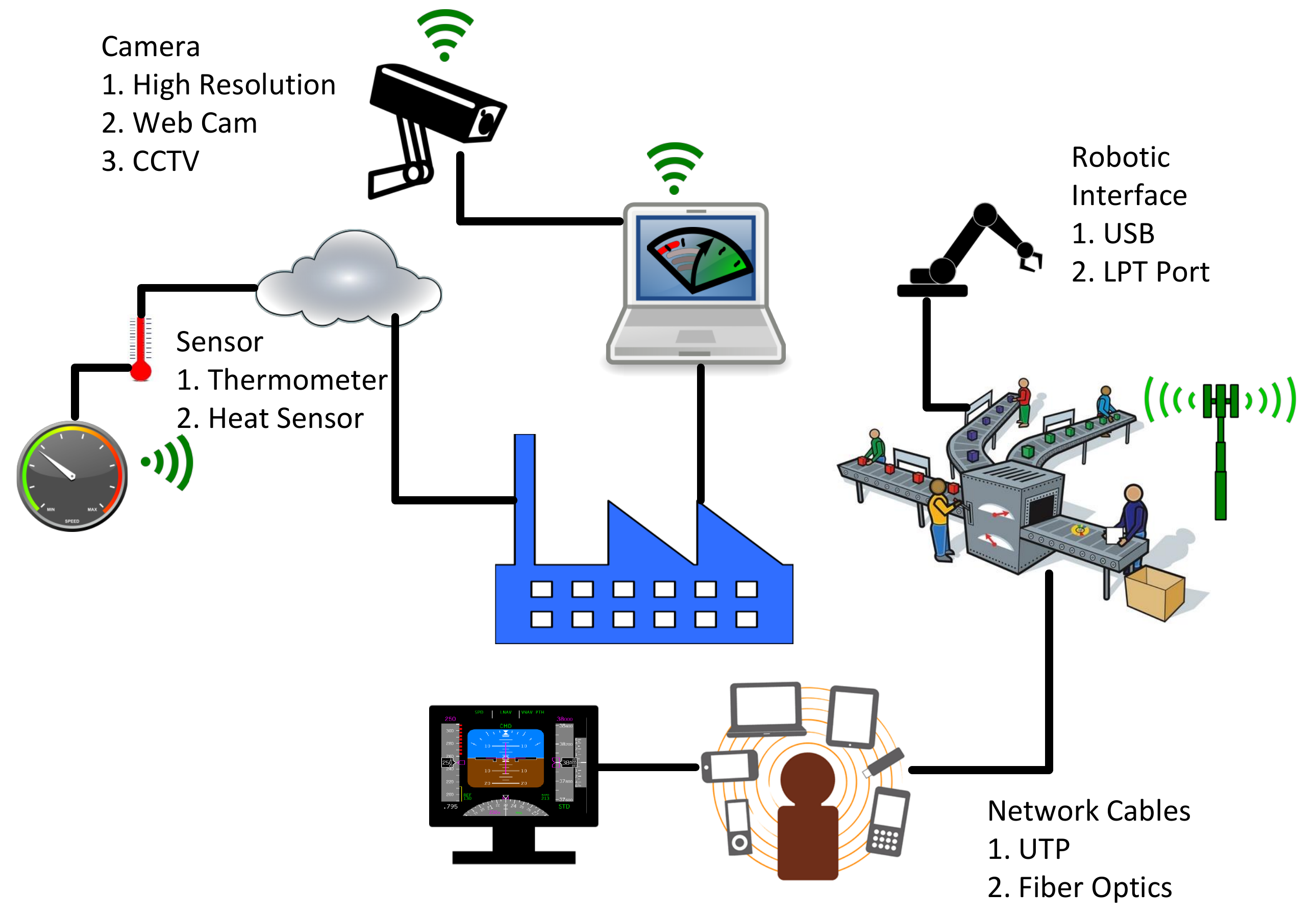}

\caption{Interconnected Manufacturing System}

\label{Figure:InterConnecteSystem}

\end{figure*}

To illustrate the use of CA for $t-wise$ testing, consider a hypothetical example of an integrated manufacturing system in Figure~\ref{Figure:InterConnecteSystem}. There are four basic elements/parameters of the system, i.e., Camera, Robotic Interface, Sensor, and Network Cables. The camera parameter takes three possible values (i.e., Camera = $\lbrace$High Resolution, Web Cam, and CCTV$\rbrace$), whereas the rest of the parameters take two possible values (i.e., Robotic Interface = $\lbrace$USB, HDMI$\rbrace$, Sensor = $\lbrace$Thermometer, Heat Sensor$\rbrace$, and Network Cables = $\lbrace$UTP, Fiber Optics$\rbrace$).

\begin{figure*}

\centering

\includegraphics[width= 3.5 in]					{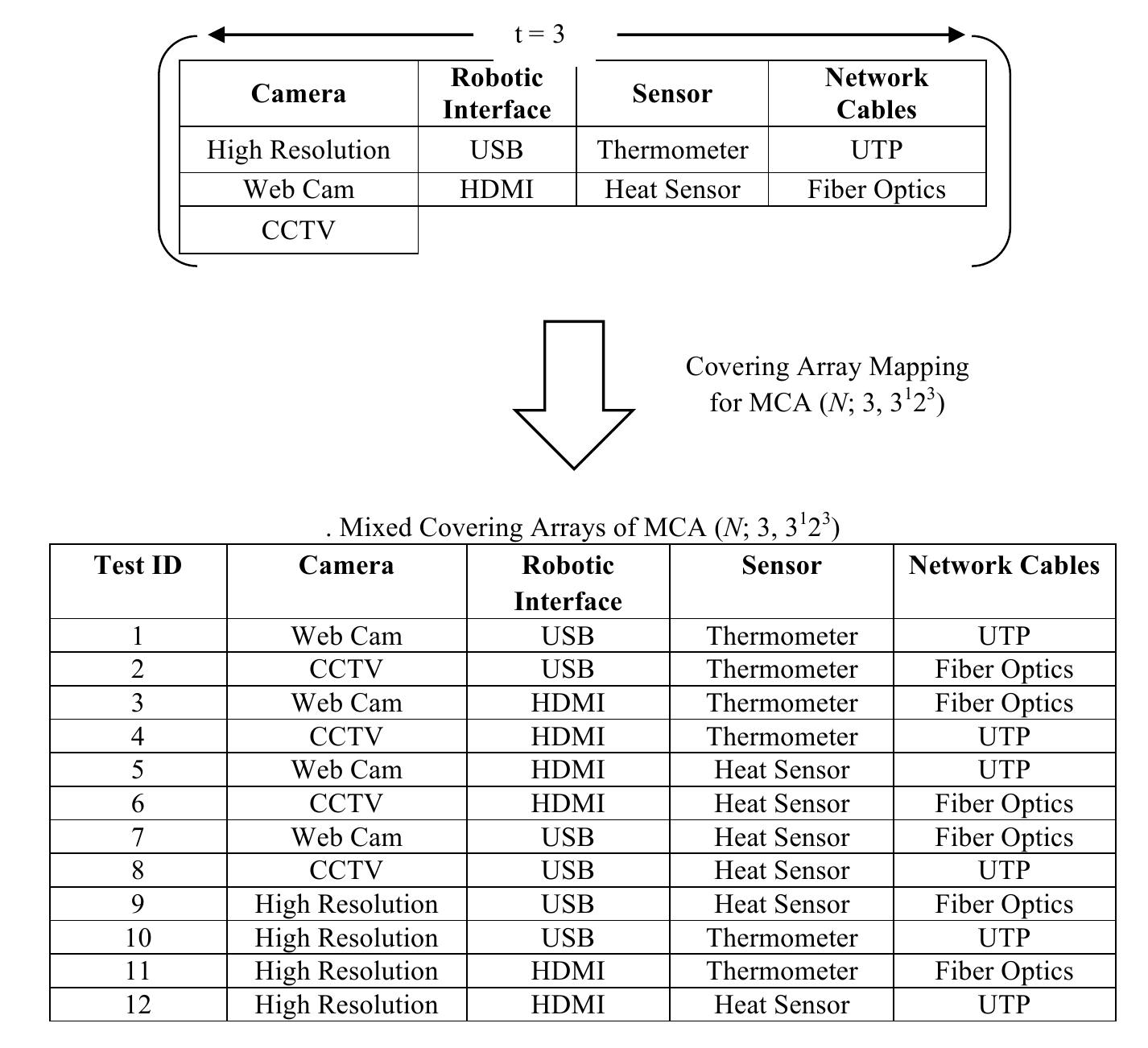}

\caption{Mixed CA Construction $MCA (N; 3,3^1, 2^3)$ for
Interconnected Manufacturing System
}

\label{Fig2:InterConnecteSystem}

\end{figure*}

As an example, the mixed CA representation for $MCA (N; 3, 3^1 2^3)$ is shown in Figure~\ref{Fig2:InterConnecteSystem} with twelve test cases. In this case, there is a reduction of 50\% test cases from the 24 exhaustive possibilities.

\section{Related Work}
In this section, we present the previous work performed on the combinatorial $t-wise$ test generation and the evaluation of such techniques in terms of efficiency and effectiveness. 

\subsection{Combinatorial $t-wise$ test suite generators}
CA construction is an NP-complete problem~\cite{IpoNP1998}. CA construction is directly applied for $t-wise$ test case reduction; thus, considerable research has been carried out to develop effective strategies for obtaining (near) optimal solutions. Existing works for CA generation can be classified into two main approaches, namely mathematical and greedy computational approaches. The mathematical approach often exploits the mathematical properties of orthogonal arrays to construct efficient CA~\cite{Mandl1985}. An example of strategies that originate from the extension of mathematical concepts called orthogonal array is recursive CA~\cite{JCD20065}. The main limitation of the OA solutions is that these techniques restrict the selection of values, which are confined to low interaction (i.e., $t < 3$), thus, limiting its applicability for only small-scale systems configurations. Greedy computational approaches exploit computing power to generate the required CA, such that each solution results from the greedy selection of the required interaction. The greedy computational approaches can be categorized further into one-parameter-at-a-time (OPAT) and one-test-at-a-time (OTAT) methods~\cite{Nie2011}. In-parameter-order (IPO) strategy~\cite{IpoNP1998} is perhaps the pioneer strategy that adopts the OPAT approach (hence termed IPO-like). IPO strategy is later generalized into a number of variants IPOG~\cite{LeiIPOG2007}, IPOG-D~\cite{Lei2008IET}, IPOF~\cite{Forbes2008}, and IPO-s~\cite{Calvagna2009}.Whereas, AETG~\cite{Cohen605761} is the first CA construction strategy that adopts the OTAT method (hence, termed AETG-like~\cite{Williams558835}). Many variants of AETG emerged later, including mAETG~\cite{Cohen04designingtest}, and $mAETG_SAT$~\cite{Cohen:2007}.

One can find two recent trends in research for combinatorial interaction testing: handling of constraints~\cite{8102999} and the application of meta-heuristic algorithms. Many current studies focus on the use of meta-heuristic algorithms as part of the greedy computational approach for CA construction~\cite{Mahmoud:2015,Wu6919298,Ahmed:2012}. Meta-heuristic-based strategies, which complement both the OPAT and OTAT methods, are often superior in terms of obtaining optimal CA size, but trade-offs regarding computational costs may exist. Meta-heuristic-based strategies often start with a population of random solutions. One or more search operators are iteratively applied to the population to improve the overall fitness (i.e., regarding greedily covering the interaction combinations). Although variations are numerous, the main difference between meta-heuristic strategies is on the defined search operators.  Meta-heuristics such as genetic algorithm (e.g. GA)~\cite{Shiba:2004}, ant colony optimization (e.g. ACO)~\cite{Chen:2009}, simulated annealing (e.g. SA)~\cite{Cohen:2007}, particle swarm optimization (e.g. PSTG~\cite{Ahmed:2012}, DPSO)~\cite{Wu6919298}, and cuckoo search algorithm (e.g. CS)~\cite{AhmedBestoun:2015} are effectively used for CA construction.

In line with the development of meta-heuristic algorithms, the room for improvement is substantial to advance the field of Search-Based Software Engineering (SBSE)  by the provision of hybridizing two or more algorithms. Each algorithm usually has its advantages and disadvantages. With hybridization, each algorithm can exploit the strengths and cover the weaknesses of the collaborating algorithms (i.e., either partly or in full). Many recent scientific results indicate that hybridization improves the performance of meta-heuristic algorithms~\cite{SABAR2015225}.

Owing to its ability to accommodate two or more search operators from different meta-heuristics (partly or in full) through one defined parent heuristic \cite{Burke2013}, hyper-heuristics can be seen as an elegant way to support hybridization. To be specific, the selection of a particular search operator at any particular instance can be adaptively decided (by the parent meta-heuristic) based on the feedback from its previous performance (i.e., learning).

In general, hyper-heuristic can be categorized as either selective or generative ones \cite{Burke2010-1}. Ideally, a selective hyper-heuristic can select the appropriate heuristics from a pool of possible heuristics. On the other hand, a generative hyper-heuristic can generate new heuristics from existing ones. Typically, selective and generative hyper-heuristics can be further categorized as either constructive or perturbative ones. A constructive gradually builds a particular solution from scratch. On the other hand, a perturbative hyper-heuristic iteratively improves an existing solution by relying on its perturbative mechanisms.

In hyper-heuristic, there is a need to maintain a “domain barrier” that controls and filters out domain-specific information from the hyper-heuristic itself \cite{Burke2013-2}. In other words, hyper-heuristic ensures generality to its approach.

Concerning related work for CA construction, Zamli \textit{et al.}\cite{Zamli:2016:TSH} implemented tabu search hyper-heuristic (Tabu HHH) utilizing a selection hyper-heuristic based on tabu search and three measures (quality, diversify and intensify) to assist the heuristic selection process. Although showing promising results, Tabu HHH adopted full meta-heuristic algorithms (i.e., comprising of Teaching Learning based Optimization (TLBO) \cite{Rao:2011:TON}, Particle Swarm Optimization (PSO) \cite{Kennedy488968}, and Cuckoo Search Algorithm (CS) \cite{Suash2009}) as its search operators. Using the three measures in HHH, Zamli \textit{et al.}~\cite{Zamli:2017:ESH} later introduced the new Mamdani fuzzy based hyper-heuristic that can accommodate partial truth, hence, allowing a smoother transition between the search operators. In other work, Jia \textit{et al.}~\cite{Cohen7194604} implemented a simulated annealing-based hyper-heuristic called HHSA to select from variants of six operators (i.e., single/multiple/smart mutation, simple/smart add and delete row). HHSA demonstrates good performance regarding test suite size and exhibits elements of learning in the selection of the search operator.
 
Complementing HHSA, we propose Q-EMCQ as another alternative SA variant. Unlike HHSA, we integrate the Q-learning mechanism to provide a memory of the performance of each search operator for selection. The Q-learning mechanism complements the Monte-Carlo based exponential Metropolis probability function by keeping track of the best performing operators for selection when the current fitness function is poor. Also, unlike HHSA, which deals only with CA (with constraints) construction, our work also focuses on MCA.

\subsection{Case studies on combinatorial $t-wise$ interaction test generation}
The number of successful applications of combinatorial interaction testing in the literature is expanding. Few studies~\cite{kuhn2006pseudo,kuhn2004software,bell2005effectiveness,wallace2001failure,charbachi2017can,bergstrom2017using,sampath2012improving,charbachi2017can} are focusing on fault and failure detection capabilities of these techniques for different industrial systems. However, still, there is a lack of industrial applicability of combinatorial interaction testing strategies.

Some case studies concerning combinatorial testing have focused on comparing between different strengths of combinatorial criteria~\cite{grindal2006evaluation} with random tests~\cite{ghandehari2014empirical,schroeder2004comparing} and the coverage achieved by such test cases. For example, Cohen \textit{et al.}~\cite{cohen1996combinatorial} found that pairwise generated tests can achieve 90\% code coverage by using the AETG tool. Other studies~\cite{cohen1994automatic,dalal1998model,sampath2012improving} have reported the use of combinatorial testing on real-world systems and how it can help in the detection of faults when compared to other test design techniques. 

Few papers examine the effectiveness (i.e., the ability of test cases to detect faults) of combinatorial tests of different $t-wise$ strengths and how these strategies compare with each other. There is some empirical evidence suggesting that across a variety of domains, all failures could be triggered by a maximum of $4-way$ interactions~\cite {kuhn2006pseudo,kuhn2004software,bell2005effectiveness,wallace2001failure}. In one such case, 67\% of failures are caused by one parameter, $2-way$ combinations cause 93\% of failures, and 98\% by $3-way$ combinations. The detection rate for other studies is similar, reaching 100\% fault detection by the use of $4-way$ interactions. These results encouraged our interest in investigating a larger case study on how Q-EMCQ and different interaction strengths perform in terms of test efficiency and effectiveness for industrial software systems and study the degree of interaction involved in detecting faults for such programs.

\section{Overview of the Proposed Strategy}

\begin{figure*}
\centering
\includegraphics[width=3.8 in]                {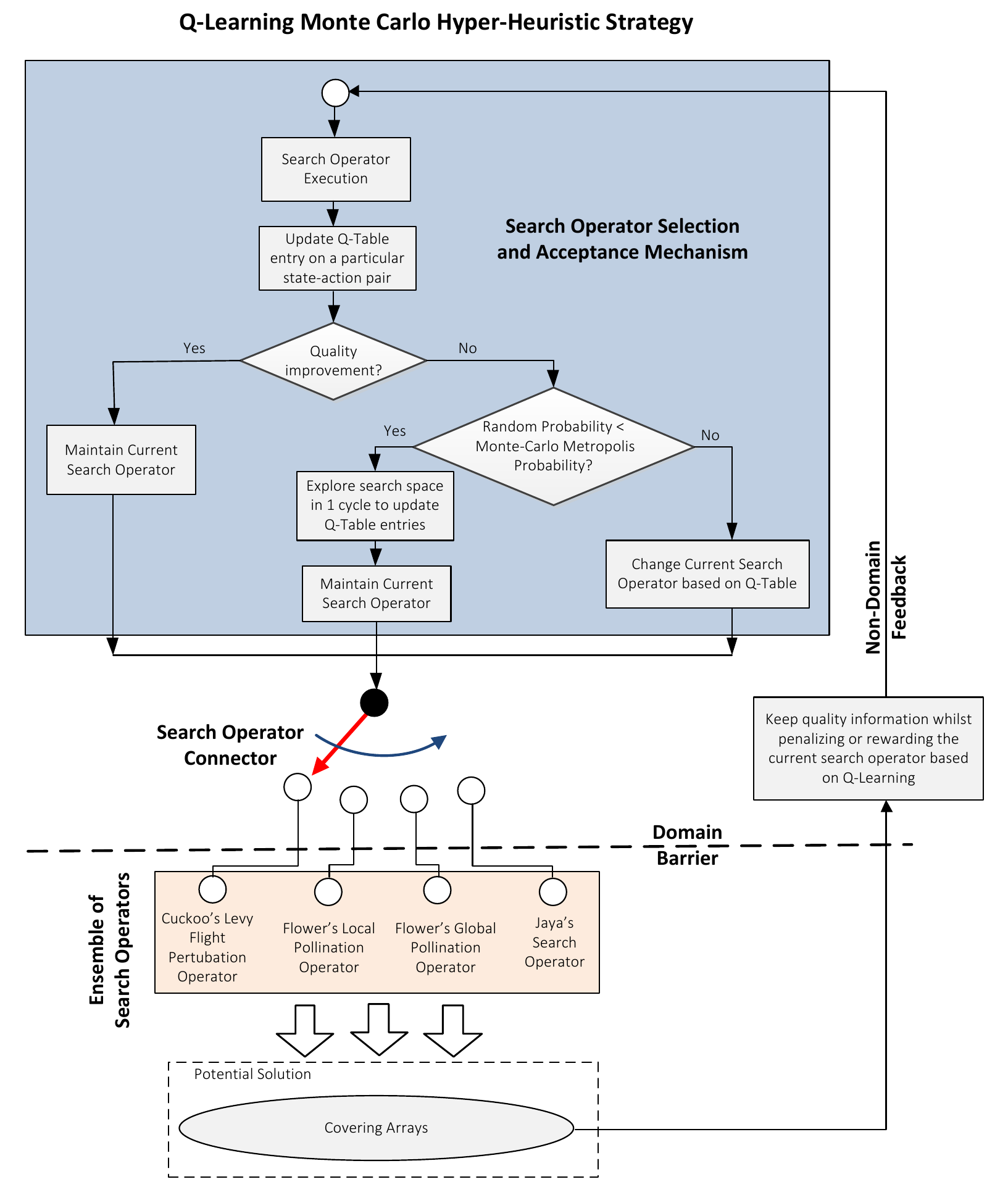}
\caption{High Level View of the Proposed Hyper-Heuristic Strategy}
\label{Fig:HyperGraph}
\end{figure*}

The high-level view of Q-EMCQ strategy is illustrated in Figure~\ref{Fig:HyperGraph}. The main components of Q-EMCQ consist of the algorithm (along with its selection and acceptance mechanism) and the defined search operators. Referring to Figure 3, Q-EMCQ chooses the search operator much like a multiplexer via a search operator connector based on the memory on its previous performances (i.e., penalize and reward). However, it should be noted that the Q-learning mechanism is only summoned when there are no improvements in the prior iteration. The complete and details working of Q-EMCQ is highlighted in the next subsections.

\subsection{Q-Learning Monte Carlo Hyper-Heuristic Strategy}

The exponential Monte Carlo with counter (EMCQ) algorithm from~\cite{Ayob03amonte,Batat2014} has been adopted in this work as the basis of Q-EMCQ selection and acceptance mechanism. EMCQ algorithm accepts poor solution (similar to simulated annealing~\cite{Kirkpatrick671}, the probability density is defined as: 

\begin{equation}
\psi  = {e^{ - \frac{{\delta T}}{q}}}
\end{equation}

where $\delta$ is the difference in fitness value between the current solution ($S_{i}$) and the previous solution ($S_{0}$) (i.e. $\delta=f(S_{i})-f(S_{0})$), $T$ is the iteration counter, and $q$ is a control parameter for consecutive non-improving iterations.

Similar to simulated annealing, probability density $\varPsi$ decreases toward zero as $T$ increases. However, unlike simulated annealing, EMCQ does not use any specific cooling schedule; hence, specific parameters do not need to be tuned. Another notable feature is that EMCQ allows dynamic manipulation on its \textit{q} parameter to increase or decrease the probability of accepting poor moves. $q$ is always incremented upon a poor move and reset to 1 upon a good move to enhance the diversification of the solution.

Although adopting the same cooling schedule as EMCQ, Q-EMCQ has a different reward and punishment mechanism. For EMCQ, the reward is based solely on the previous performance (although sometimes the poor performing operator may also be rewarded based on some probability). Unlike EMCQ, when a poor search operator is penalized, Q-EMCQ chooses the historically best performing operator for the next search instance instead of from the available search operators randomly.

Q-learning is a Markov decision process that relies on the current and forward-looking Q-values. It provides the reward and punishment mechanism~\cite{Watkins1992} that dynamically keeps track of the best performing operator via online reinforcement learning. To be specific, Q-learning learns the optimal selection policy by its interaction with the environment. Q-learning works by estimating the best state-action pair through the manipulation of memory based on $Q(s, a)$ table. A $Q(s,a)$ table uses a state-action pair to index a $Q$ value (i.e. as cumulative reward). The $Q(s, a)$ table is updated dynamically based on the reward and punishment $(r)$ from a particular state-action pair.

Let $S=[s_{1},s_{2},\ldots,s_{n}]$ be a set of states, $A=[a_{1},a_{2},\ldots,a_{n}]$ be a set of actions, $\alpha_{t}$ be the learning rate within $[0,1]$, $\gamma$ be the discount factor within $[0,1]$, $r_{t}$ be the immediate reward/punishment acquired from executing action a, the $Q(st,at)$ as the cumulative reward at time $(t$) can be computed as follows:

\begin{equation}
\begin{split}
Q_{(t+1)}(s_{t},a_{t})=Q_{t}(s_{t},a_{t})+\alpha_{t}(r_{t}+\gamma max(Q_{t}\\(s_{(t+1)},a_{(t+1)}))-Q_{t}(s_{t},a_{t}))
\label{eq2}
\end{split}
\end{equation}

The optimal setting for $t$, $\gamma$, and $r_{t}$ needs further clarification. When $\alpha_{t}$ is close to 1, a higher priority is given to the newly gained information for the Q-table updates. On the contrary, a small value of $\alpha_{t}$ gives higher priority to existing information. To facilitate exploration of the search space (to maximize learning from the environment), the value of $\alpha_{t}$ during early iteration can be set a high value, but adaptively reduce towards the end of the iteration (to exploit the existing best known Q-value) as follows:

\begin{equation}
\alpha_{t}=1-0.9\times t/(MaxIteration)
\end{equation}

The parameter $\gamma$ works as the scaling factor for rewarding or punishing the Q-value based on the current action. When $\gamma$ is close to 0, the Q-value is based on the current reward/punishment only. When $\gamma$ is close to 1, the Q-value will be based on the current and the previous reward/punishment. It is suggested to set $\gamma$= 0.8~\cite{Samma:2016}.

The parameter $r_t$ serves as the actual reward or punishment value. In our current work, the value of $r_t$ is set based on:

\begin{equation}
\left. \begin{array}{l}
{r_t} = 1,\;if\;the\;current\; action\;improves\;fitness\\
{r_t} =  - 1,\;otherwise
\end{array} \right\}
\end{equation}

Based on the discussion above, Algorithm~\ref{Pseudo_Q-EMCQ} highlights the pseudo-code for Q-EMCQ.

\begin{algorithm*}

\caption{Pseudo Code for Q-EMCQ}
\label{Pseudo_Q-EMCQ}
\includegraphics[width=3.9 in]                {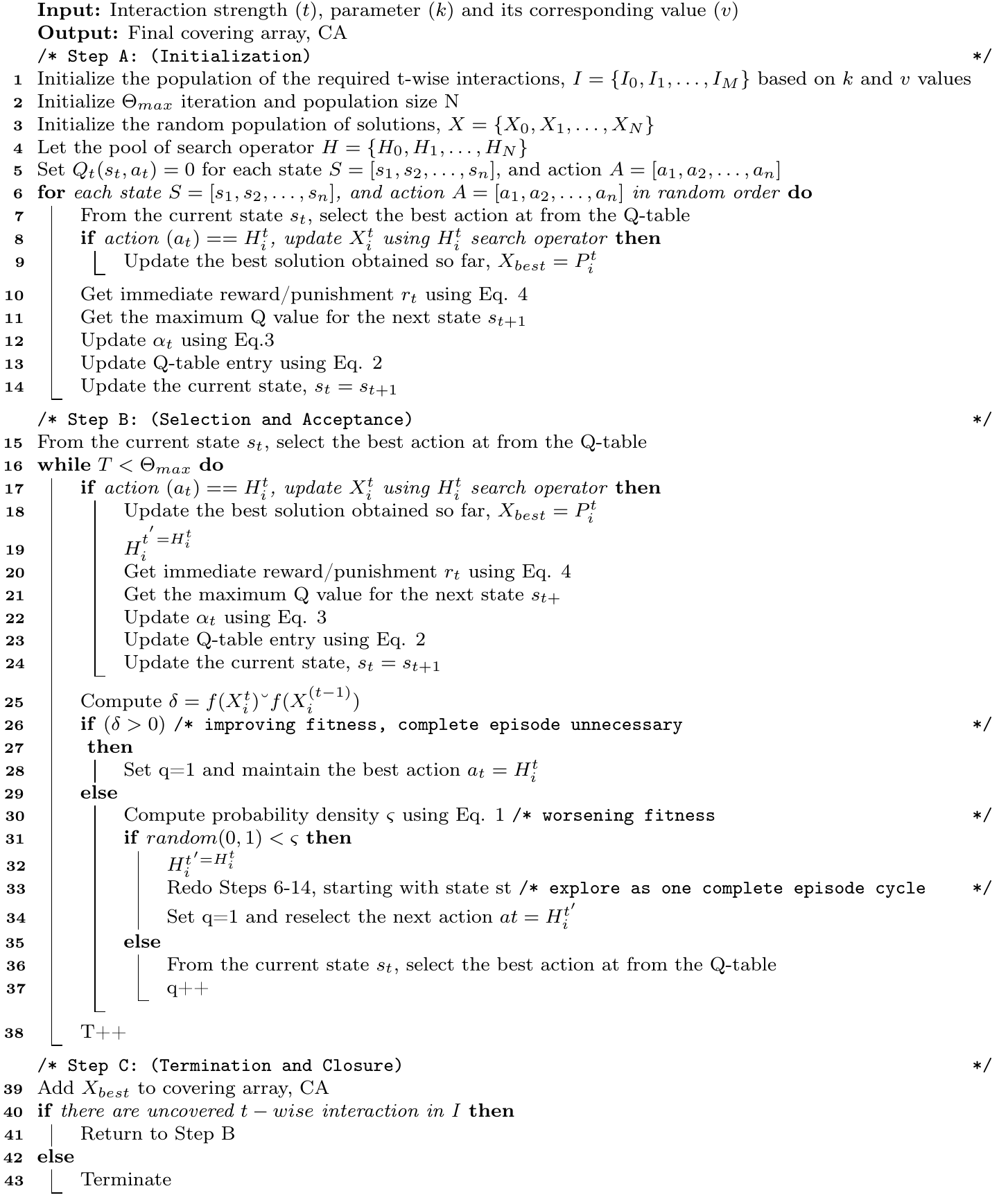}

\end{algorithm*}

Q-EMCQ involves three main steps, denoted as Steps A, B, and C. Step A deals with the initialization of variables. Line 1 initializes the populations of the required $t-wise$ interactions, $I={I_{1},I_{2},\ldots,I_{M}}$. The value of $M$ depends on the given inputs interaction strength ($t$), parameter ($k$), and its corresponding value ($v$). $M$ captures the number of required interactions that needs to be captured in the constructed CA. $M$ can be mathematically obtained as the sum of products of each individual\textquoteright s $t-wise$ interaction. For example, for $CA(9;2,3^{4})$, $M$ takes the value of  $3\times3+3\times3+3\times3+3\times3+3\times3+3\times3=54$. If $MCA(9;2,3^{2}2^{2})$ is considered, then $M$ takes the value of $3\times3+3\times2+3\times2+3\times2+3\times2+2\times2=37$. Line 2 defines the maximum iteration $\Theta_{max}$ and population size, $N$. Line 3 randomly initializes the initial population of solution $X={X_{1},X_{2},\ldots,X_{M}}$. Line 4 defines the pool of search operators. Lines 6-14 explore the search space for 1 complete episode cycle to initialize the Q-table.

\begin{figure*}

\centering

\includegraphics[width= 5.4 in]                    {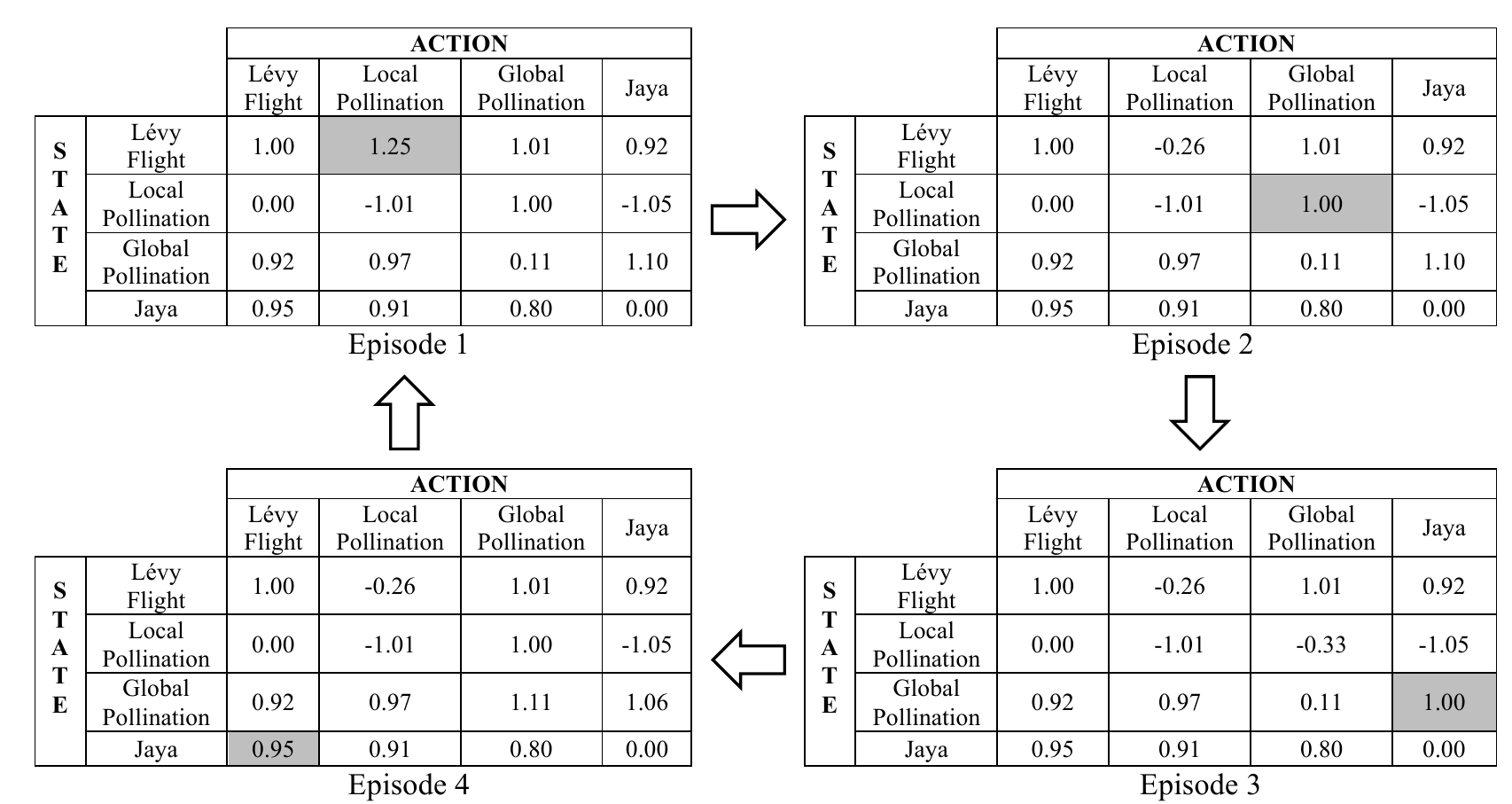}

\caption{Q-Learning Mechanism for 1 Complete Episode Cycle}
\label{Fig:Q-Learning-Mechanism}

\end{figure*}

Step B deals with the Q-EMCQ selection and acceptance mechanism. The main loop starts in line 15 with $\Theta_{max}$ as the maximum number of iteration. The selected search operator will be executed in line 17. The Q-table will be updated accordingly based on the quality/performance of the current state-action pairs (lines 18-24). Like EMCQ, the Monte Carlo Metropolis probability controls the selection of search operators when the quality of the solution improves (lines 25-30). This probability decreases with iteration ($T$). However, it may also increase as the $q$ value can be reset to 1 (in the case of re-selection of any particular search operator (lines 29 and 34). When the quality does not improve, the $Q$ learning gets a chance to explore the search space in one complete episode cycle (as line 33) to complete the Q-table entries. As an illustration, Figure~\ref{Fig:Q-Learning-Mechanism} depicts the snapshot of one entire Q-table cycle for Q-EMCQ along with a numerical example.

Referring to episode 1 in Figure~\ref{Fig:Q-Learning-Mechanism}, assume that the initial settings are as follows: the current state $s_{t}$= \textit{Lévy Flight Perturbation Operator}, the next action $a_{t}$= \textit{Local Pollination Operator}, the current value stored in the Q-table for the current state $Q_{(t+1)}(s_{t},a_{t})=1.25$ (i.e. greyed cell); the punishment $r_{t}=-1.00$; the discount factor $\gamma=0.10$; and the current learning factor $\alpha_{t}=0.70$.
Then the new value for $Q_{(t+1)}(s_{t},a_{t})$ in the Q-table is updated based on Eq.~\ref{eq2} as:

\begin{equation}
\begin{split}
Q_{(t+1)}(s_{t},a_{t})=1.25+0.70\times[-1.00+0.10\times\\ Max(0.00,-1.01,1.00,-1.05)-1.25]=-0.26
\end{split}
\end{equation}

Concerning episode 2 in Figure~\ref{Fig:Q-Learning-Mechanism}, the current settings are as follows: the current state $s_{t}$= \textit{Local Pollination Operator}, the next action $a_{t}$= \textit{Global Pollination Operator}, the current value stored in the Q-table for the current state $Q_{(t+1)}(s_{t},a_{t})=1.00$ (i.e. greyed cell ); the punishment $r_{t}=-1.00$; the discount factor $\gamma=0.10$; and the current learning factor $\alpha_{t}=0.70$. Then the new value for $Q_{(t+1)}(s_{t},a_{t})$ in the Q-table is updated based on Eq.~\ref{eq2} as:

\begin{equation}
\begin{split}
Q_{(t+1)}(s_{t},a_{t})=1.00+0.70\times[-1.00+0.10\times\\ Max(0.92,0.97,0.11,1.00)-1.00]=-0.33
\end{split}
\end{equation}

Considering episode 3 in Figure~\ref{Fig:Q-Learning-Mechanism}, the current settings are as follows: the current state $s_{t}=$ \textit{Global Pollination Operator}, the next action $a_{t}=$ \textit{Jaya Operator}, the current value stored in the Q-table for the current state $Q_{(t+1)}(s_{t},a_{t})=1.00$ (i.e. greyed cell ); the reward $r_{t}=1.00$; the discount factor $\gamma=0.10$; and the current learning factor $\alpha t=0.70$. Then the new value for $Q_{(t+1)}(s_{t},a_{t})$ in the Q-table is updated based on Eq.~\ref{eq2} as:

\begin{equation}
\begin{split}
Q_{(t+1)}(s_{t},a_{t})=1.00+0.70\times[1.00+0.10\times\\ Max(0.95,0.91,0.80,0.00)-1.00]=1.06
\end{split}
\end{equation}

The complete exploration cycle for updating Q-values ends in episode 4 as the next action $a_{t}=s_{(t+1)}=$ \textit{Lévy Flight Perturbation Operator}. It must be noted that throughout the Q-table updates, the Q-EMCQ search process is also working in the background (i.e., for each update, $X_{best}$ is also kept and the population $X$ is also updated accordingly).

A complete cycle update is not always necessary, especially during convergence. Lines 38-39 depicts the search operator selection process as the next action ($a_{t}$) (i.e. between Lévy Flight Perturbation Operator, Local Pollination Operator, Global Pollination Operator, and Jaya Operator) based on the maximum reward defined in the state-action pair memory within the Q-table (unlike EMCQ where the selection process is random).

Complementing earlier steps, Step C deals with termination and closure. In line 39, upon the completion of the main $\Theta_{max}$ loop, the best solution $S_{best}$ is added to the final CA. If uncovered $t-wise$ interaction exists, Step B is repeated until termination (line 41).

\subsection{Cuckoo’s Levy Flight Perturbation Operator}

Cuckoo’s Levy flight perturbation operator is derived from the cuckoo search algorithm (CS)~\cite{Suash2009}. The complete description of the perturbation operator is summarized in Algorithm~\ref{PseudoCuckoo_Levy}.

Cuckoo’s Levy flight perturbation operator acts as the local search algorithm that manipulates the \textit{Lévy flight} motion. For our \textit{Lévy flight} implementation, we adopt the well-known Mantegna’s algorithm~\cite{Suash2009}. Within this algorithm, a \textit{Lévy flight} step length can be defined as:

\begin{equation}
Step=u/[v]^{(1/\beta)}
\end{equation}

where $u$ and $v$ are approximated from the normal Gaussian distribution
in which

\begin{equation}
u\text{\ensuremath{\approx}}N(0,\sigma_{u}{}^{2})\times\sigma_{u\;\quad\quad}v\text{\ensuremath{\approx}}N(0,\sigma_{v}{}^{2})\times\sigma_{v}
\end{equation}

For $v$ value estimation, we use $\sigma_{v}=1$. For $u$ value estimation, we evaluate the Gamma function ($\Gamma$) with the value of $\beta=1.5$ \cite{Yang:2008:NMA}, and obtain $\sigma_{u}$ using

\begin{equation}
\sigma_{u}=|\frac{(\Gamma(1+\beta)\times sin(\pi\beta/2))}{(\Gamma(1+\beta)/2)\times\beta\times2^{(((\beta-1))/2)})}|^{(1/\beta)}
\end{equation}

In our case, the Gamma function ($\Gamma$) implementation is adopted from William \textit{et al.} \cite{Press:1992:NRC}. The Lévy flight motion is essentially a random walk that takes a sequence of jumps, which are selected from a heavy-tailed probability function \cite{Suash2009}. As a result, the motion will produce a series of \textquotedblleft aggressive\textquotedblright{} small and large jumps (either positive or negative), thus ensuring largely diverse values. In our implementation, the Lévy flight motion performs a single value perturbation of the current population of solutions, thus rendering it as a local search operator.

\begin{algorithm*}

\caption{Pseudo Code for Cuckoo’s Levy Flight Perturbation Operator}
\label{PseudoCuckoo_Levy}
\scriptsize
\KwIn{the population $X = \lbrace{X_0, X_1, \dots , X_{M} \rbrace}$}
 \KwOut{$X_{best}$ and the updated population $X^{'} = \lbrace{X_{0}^{’}, X_{1}^{’}, \dots , X_{M}^{’} \rbrace} $}
 
 $X_{best} = X_0$
 
 \For{$i=0$ to population size, $M$}{
 
 Generate a step vector $\L$ which obeys Levy Flight distribution
 
 Perturbate one value from random column wise,  $X_{i}^{t+1} = X_{i}^{t}  + \alpha \bigoplus \L \, with \, \alpha =1$
 
 \eIf{$f(X_i^{(t+1)}) > f(X_i^{(t)})$ }{
 
 $X_i^{(t)} = X_i^{(t+1)} $
 
 \If{($f(X_i^{(t+1)}) > f(X_{best}))$}{
 
 $X_{best}= X_i^{(t+1)}$
 }
 }
 {
 
 \If{($f(X_i^{(t)}) > f(X_{best})$}{
 
 $X_{best}= X_i^{(t)}$
 }

 }

 }

Return $S_{best}$

\end{algorithm*}

As for the working of the operator, the initial $X_{best}$ is set to $X_{0}$ in line 1. The loop starts on line 2. One value from a particular individual $X_{i}$ is selected randomly (column-wise) and perturbed using $\alpha$ with entry-wise multiplication ($\oplus$) and levy flight motion ($L$), as indicated in line 4. If the newly perturbed $X_{i}$ has a better fitness value, then the incumbent is replaced and the value of $X_{best}$ is also updated accordingly (in lines 5\textendash 11). Otherwise, $X_{i}$ is not updated, but $X_{best}$ will be updated based on its fitness against $X_{i}$.

\subsection{Flower’s Local Pollination Operator}

As the name suggests, the flower’s local pollination operator is derived from the flower algorithm~\cite{YangShi2012}. The complete description of the operator is summarized in Algorithm~\ref{AlgoFlowerLocal}.

\begin{algorithm*}

\caption{Flower’s Local Pollination Operator}
\label{AlgoFlowerLocal}
\scriptsize

  \KwIn{the population $X = \lbrace{X_0, X_1, \dots , X_{M} \rbrace}$}
 \KwOut{$X_{best}$ and the updated population $X^{'} = \lbrace{X_{0}^{’}, X_{1}^{’}, \dots , X_{M}^{’} \rbrace} $}
 
 $X_{best} = X_0$

\For{$i=0$ to population size, $S-1$}{
 
Choose $X_p$ and $X_q$ randomly from X, where $j \neq k$

Set $\gamma = random \, (0,1)$

Update the current population $X_i^{(t+1)} = X_i^{(t)} + \gamma (X_p^{(t)} – X_q^{(t)})$
 
\eIf{($f(X_i^{(t+1)}) > f(X_i^{(t)})$)}{

$X_i^{(t)} = X_i^{(t+1)} $

\If{$(f(X_i^{(t+1)}) > f(X_{best}))$}{

$X_{best}= X_i^{(t+1)}$
}

}
{
\If{\textbf{$(f(X_i^{(t)}) > f(X_{best}))$}}{

$X_{best}= X_i^{(t)}$

}
}

 }

Return $S_{best}$

\end{algorithm*}

In line 1, $XS_{best}$ is initially set to $X_{0}$. In line 2, two distinct peer candidates $X_{p}$ and $X_{q}$ are randomly selected from the current population $X$. The loop starts on line 2. Each $X_{i}$ will be iteratively updated based on the transformation equation defined in lines 4\textendash 5. If the newly updated $X_{i}$ has better fitness value, then the current $X_{i}$ is replaced accordingly (in lines 6-7). The value of $X_{best}$ is also updated if it has a better fitness value than that of $X_{i}$ (in lines 8\textendash 10). When the newly updated $X_{i}$ has poorer fitness value, no update is made to $X_{i}$, but $X_{best}$ will be updated if it has better fitness than $X_{i}$ (in lines 11\textendash 12).

\subsection{Flower’s Global Pollination Operator}

Flower’s global pollination operator~\cite{YangShi2012} is summarized in Algorithm~\ref{AlgoFlowerGlobal} and complements the local pollination operator described earlier.

\begin{algorithm*}

\caption{Flower’s Global Pollination Operator}
\label{AlgoFlowerGlobal}
\scriptsize

 \KwIn{the population $X = \lbrace{X_0, X_1, \dots , X_{M} \rbrace}$}
 \KwOut{$X_{best}$ and the updated population $X^{'} = \lbrace{X_{0}^{’}, X_{1}^{’}, \dots , X_{M}^{’} \rbrace} $}
 
 $X_{best} = X_0$

 \For{$i=0$ to population size, $M$}{
 
Set scaling factor $\rho  = random(0,1)$
 
Generate a step vector $\L$ which obeys Levy Flight distribution
 
Update the current population $X_i^{(t+1)} = X_i^{(t)} + \rho \cdot \L \cdot (X_{best} – X_i^{(t)})$

\eIf{$(f(X_i^{(t+1)}) > f(X_i^{(t)})) $}{

$X_i^{(t)} = X_i^{(t+1)} $

\If{$(f(X_i^{(t+1)}) > f(X_{best}))$}{

$X_{best}= X_i^{(t+1)}$
}

}{

\If{$(f(X_i^{(t)}) > f(X_{best}))$}{

$X_{best} = X_i^{(t)}$
}

}

 }

Return $X_{best}$

\end{algorithm*}

Similar to cuckoo\textquoteright s Levy flight perturbation operator described earlier, the global pollination operator also exploits Levy flight motion to generate a new solution. Unlike the former operator, the transformation equation for flower\textquoteright s global pollination operator uses the Levy flight to update all the (column-wise) values for $Z_{i}$ of interest instead of only perturbing one value, thereby making it a global search operator.

Considering the flow of the global pollination operator, $X_{best}$ is initially set to $X_{0}$ in line 1. The loop starts on line 2. The value of $X_{i}$ will be iteratively updated by using the transformation equation that exploits exploiting Levy flight motion (in lines 4\textendash 5). If the newly updated $X_i$ has better fitness value, then the current $X_i$ is replaced accordingly (in lines 6-7). The value of $X_{best}$ is also updated if it has a better fitness value than that of $X_{i}$ (in lines 8-10). If the newly updated $X_{i}$ has poorer fitness value, no update is made to $X_{i}$. $X_{best}$ will be updated if it has better fitness than $X_{i}$ (in lines 8\textendash 10 and lines 11-12).

\subsection{Jaya Search Operator}

The Jaya search operator is derived from the Jaya algorithm~\cite{Rao2016}. The complete description of the Jaya operator is summarized in Algorithm~\ref{JayaSearchOperator}.

\begin{algorithm*}

\caption{Jaya Search Operator}
\label{JayaSearchOperator}
\scriptsize

 \KwIn{the population $X = \lbrace{X_0, X_1, \dots , X_{M} \rbrace}$}
 \KwOut{$X_{best}$ and the updated population $X^{'} = \lbrace{X_{0}^{’}, X_{1}^{’}, \dots , X_{M}^{’} \rbrace} $}
 
 $X_{best} = X_{0}$
 
$X_{poor} =X_{best}$

 \For{$i=0$ to population size, $M$}{
 
 Set $\varphi = random (0,1)$
 
 Set $\zeta = random (0,1)$
 
Update the current population $X_i^{(t+1)} = X_i^{(t)} + \varphi \cdot (X_{best} – X_i^{(t)}) - \zeta \cdot (X_{poor} – X_i^{(t)})$
 
 \eIf{$(f(X_i^{(t+1)}) > f(X_i^{(t)}) $}{
 
 $X_i^{(t)} = X_i^{(t+1)}$
 
 \If{$(f(X_i^{(t+1)}) > f(X_{best}))$}{
 
 $X_{best}= X_i^{(t+1)}$
 }
 
 }{

 \If{$(f(X_i^{(t)}) > f(X_{best}))$}{
 
 $X_{best} = X_i^{(t)}$
 }
 
 \If{$(f(X_i^{(t)}) < f(X_{poor}))$}{
 
$X_{poor}= X_i^{(t)}$

}

 }

 }

Return $X_{best}$

\end{algorithm*}

Unlike the search operators described earlier (i.e., keeping track of only $X_{best}$), the Jaya search operator keeps track of both $X_{best}$ and $X_{poor}$. As seen in line 6, the Jaya search operator exploits both $X_{best}$ and $X_{poor}$ as part of its transformation equation. Although biased toward the global search for Q-EMCQ in our application, the transformation equation can also address local search. In the case when $\Delta X=X_{best}-X_{poor}$ is sufficiently small, the transformation equation offset (in line with the term $\mho(X_{best}-X_{i})- \zeta(X_{poor}-X)$ will be insignificant relative the current location of $X_{i}$ allowing steady intensification.

As far as the flow of the Jaya operator is concerned, lines 1\textendash 2 sets up the initial values for $X_{best}=X_{0}$ and $X_{poor}=X_{best}$. The loop starts from line 3. Two random values $\mho$ and $\zeta$ are generated to compensate and scale down the delta differences between $X_{i}$ with $X_{best}$ and $X_{poor}$ in the transformation equation (in lines 4-5). If the newly updated $X_{i}$ has a better fitness value, then the current $X_{i}$ is replaced accordingly (in lines 7\textendash 8). Similarly, the value of $X_{best}$ is also updated if it has a better fitness value than that of $X_{i}$ (in lines 9\textendash 11). In the case in which the newly updated $X_{i}$ has poorer fitness value, no update is made to $X_{i}$. If the fitness of the current $X_{i}$ is better than that of $X_{best}$, $X_{best}$ is assigned to $X_{i}$ (in lines 12-13). Similarly, if the fitness of the current $X_{i}$ is poorer than that of $X_{poor}$, $X_{poor}$ is assigned to $X_{i}$ (in lines 14\textendash 15).

\section{Empirical Study Design}

We have put our strategy under extensive evaluation. The goals of the evaluation experiments are threefold: (1) to investigate how Q-EMCQ fare against its own predecessor EMCQ, (2) to benchmark Q-EMCQ against well-known strategies for $t-wise$ test suite generation, (3) to undertake the effectiveness assessment of Q-EMCQ using $t-wise$ criteria in terms of achieving branch coverage as well as revealing mutation injected faults based on real-world industrial applications, (4) to undertake the efficiency assessment of Q-EMCQ by comparing the test generation cost with manual testing, and (5) to compare the performance of Q-EMCQ with contemporary meta-heuristics and hyper-heuristics.

In line with the goals above, we focus on answering the following research questions:

\begin{itemize}
\item RQ1: In what ways does the use of Q-EMCQ improve upon EMCQ?

\item RQ2: How good is the efficiency of Q-EMCQ in terms of test suite minimization when compared to existing strategies?

\item RQ3: How good are combinatorial tests created using Q-EMCQ and $2-wise$, $3-wise$, and $4-wise$ at covering the code?

\item RQ4: How effective are the combinatorial tests created using Q-EMCQ for $2-wise$, $3-wise$, and $4-wise$ at detecting injected faults?

\item RQ5: How does Q-EMCQ with $2-wise$, $3-wise$, and $4-wise$ compare with manual testing in terms of cost?

RQ6: Apart from minimization problem (i.e., $t-wise$ test generation), is Q-EMCQ sufficiently general to solve (maximization) optimization problem (i.e. module clustering)?
\end{itemize}

 \subsection{Experimental Benchmark Set-Up}
 
We adopt an environment consisting of a machine running Windows 10, with a 2.9 GHz Intel Core i5 CPU, 16 GB 1867 MHz DDR3 RAM, and 512 GB flash storage. We set the population size of N = 20 with a maximum iteration value $\theta _{\max } = 2500$.  While such a choice of population size and maximum iterations could result in more than 50,000 fitness function evaluations, we limit our maximum fitness function evaluation to 1500 only (i.e., the Q-EMCQ stops when the fitness function evaluation reaches 1500). This is to ensure that we can have a consistent value of fitness function evaluation throughout the experiments (as each iteration can potentially trigger more than one fitness function evaluation). For statistical significance, we have executed Q-EMCQ for 20 times for each configuration and reported the best results during these runs. 
 
\subsection{Experimental Benchmark Procedures}

For RQ1, we arbitrarily select 6 combinations of covering arrays $CA(N;2,4^{2}2^{3})$, $CA(N;3,5^{2}4^{2}3^{2})$, $CA(N;4,5^{1}\\3^{2}2^{3})$, $MCA(N;2,5^{1}3^{3}2^{2})$, $MCA(N,3,6^{1}5^{1}4^{3}3^{3}2^{3})$ and $MCA(N,4,7^{1}6^{1}5^{1}4^{3}3^{3}2^{3})$. Here, the selected covering arrays span both uniform and non-uniform number of parameters. To ensure a fair comparison, we re-implement EMCQ using the same data structure and programming language (in Java) as Q-EMCQ before adopting it for covering array generation. Our EMCQ re-implementation also rides on the same low-level operators (i.e., cuckoo\textquoteright s Levy flight perturbation operator, flower algorithm\textquoteright s local pollination, and global pollination operator as well as Jaya\textquoteright s search operator). For this reason, we can fairly compare both test sizes and execution times.

For RQ2, we adopted the benchmark experiments mainly from Wu \textit{et al.}~\cite{Wu6919298}. In particular, we adopt two main experiments involving $CA(N;t,v^{7})$ with variable values $2\le v\le5$, $t$ varied up to 4 as well as $CA(N;t,3^{k})$ with variable number of parameters $3\le k\le12$, $t$ varied up to 4. We have also compared our strategy with those published results for those strategies that are not freely available to download. Parts of those strategies depend mainly on meta-heuristic algorithms, specifically HSS, PSTG, DPSO, ACO, and SA. The other part of those strategies is dependent on exact computational algorithms, specifically PICT, TVG, IPOG, and ITCH. We represent all our results in the tables where each cell represents the smallest size (marked as bold) generated by its corresponding strategy. In the case of Q-EMCQ, we also reported the average sizes to give a better indication of its efficiency. We opt for generated size comparison and not time because all of the strategies of interest are not available to us. Even if these strategies are available, their programming languages and data structure implementations are not the same renderings as an unfair execution time comparison. Often, the size comparison is absolute and is independent of the implementation language and data structure implementation.  

For answering RQ3-RQ5, we have selected a train control management system that has been in development for a couple of years. The system is a distributed control software with multiple types of software and hardware components for operation-critical and safety-related supervisory behavior of the train. The program runs on Programmable Logic Controllers (PLCs), which are commonly used as real-time controllers used in industrial domains (e.g., manufacturing and avionics); 37 industrial programs have been provided for which we applied the Q-EMCQ approach for minimizing the $t-wise$ test suite.

Concerning RQ6, we have selected three public domain class diagrams available freely in the public domains involving Credit Card Payment System (CCPS) \cite{REF-1}, Unified Inventory University ((UIU) \cite{REF-2} and Food Book (FB)\footnote{https://bit.ly/2XDPOPB} as our module case studies. Here, we have adopted the Q-EMCQ approach for maximizing the number of clusters so that we can have the best modularization quality (i.e., best clusters) for all given three systems’ class diagrams.

For evaluation purposes, we have adopted two groups of comparison. In the first group, we adopt EMCQ as well as Modified Choice Function \cite{Drake2015} and Tabu Search HHH \cite{Zamli:2016:TSH} implementations. It should be noted that all the hyper-heuristic rides on the same operators (i.e., Lévy flight, local pollination, global pollination, and Jaya). In the second group, we have decided to adopt the TLBO \cite{REF-4}, SCA \cite{REF-5} and SOS \cite{REF-6} implementations. Here, we are able to fairly compare the modularization quality as well as execution time as the data structure, language implementation and the running system environment are the same (apart from the same number of maximum fitness function evaluation). It should be noted that these algorithms (i.e., TLBO, SCA, SOS) do not have any parameter controls apart from population size and maximum iteration. Hence, their adoption does not require any parameter calibrations.

\subsection{Case Study Object}

As highlighted earlier, we adopt two case study objects involving the train control management system as well as the module clustering of class diagrams.

\subsubsection{Train Control Management System}

\begin{figure*}[tbp]
  	\centering
    \includegraphics[width=0.70\textwidth]{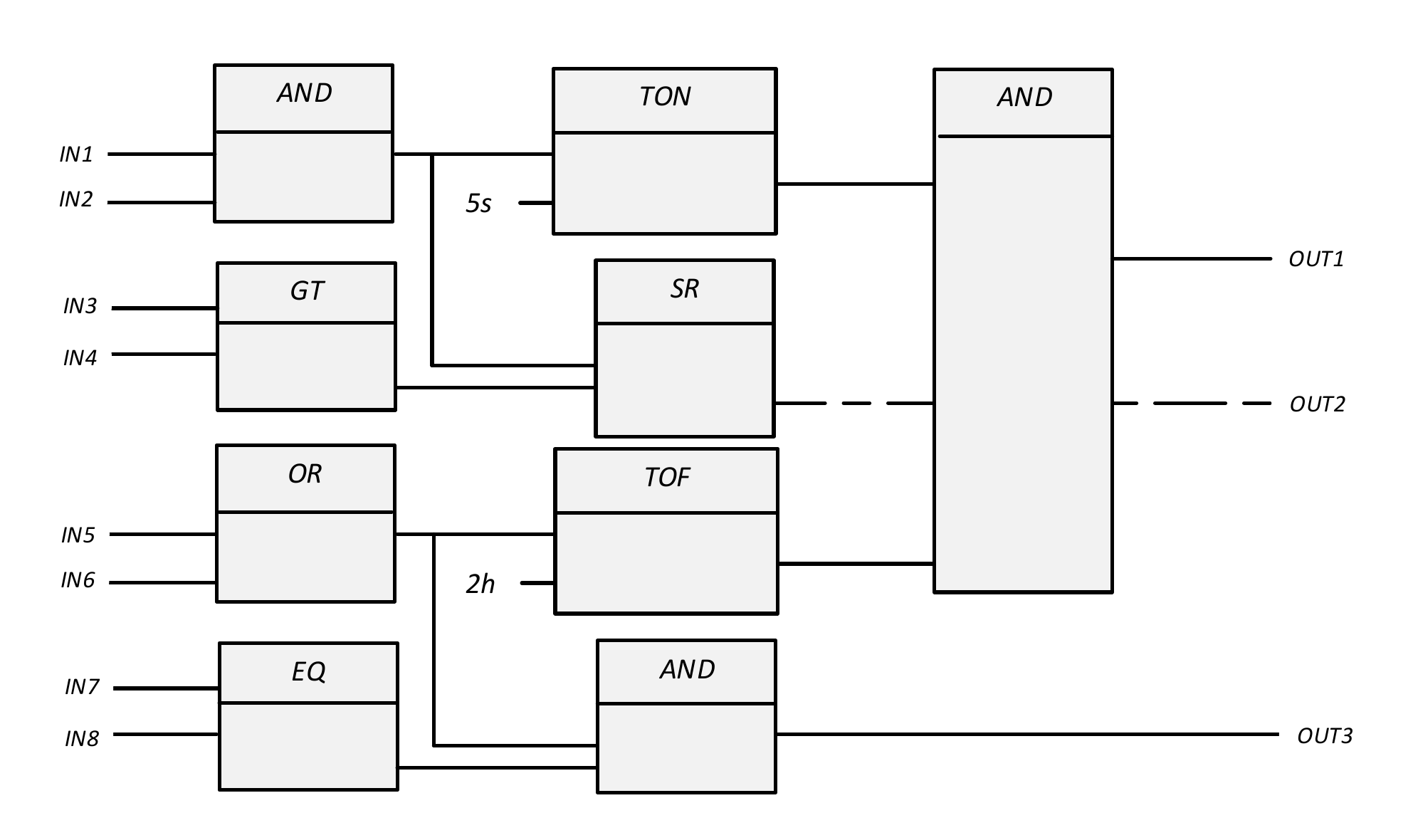}
  \caption{An example of a PLC control program written using the FBD programming language.}\label{fig:FBDtesting}
\end{figure*}

We have conducted our experiment on programs from a train control management system running on PLCs that have been developed for a couple of years. A program running on a PLC executes in a loop in which every cycle contains the reading of input values, the execution of the program without interruptions and the update of the output variables. As shown in Figure~\ref{fig:FBDtesting}, predefined logical and/or stateful blocks (e.g., bistable latch SR, OR, XOR, AND, greater-than GT and timer TON) and connections between blocks represent the behavior of a PLC program written in the Function Block Diagram (FBD) programming language \cite{john2010iec}. A hardware manufacturer supplies these blocks or is developed using custom functions. PLCs contain particular types of blocks, such as timers (e.g., TON) that provide the same functions as timing relays and are used to activate or deactivate a device after a preset interval of time. There are two different timer blocks (i) On-delay Timer (TON) and (ii) Off-delay Timer (TOF). A timer block keeps track of the number of times its input is either true or false and outputs different signals. In practice, many other timing configurations can be derived from these basic timers. An FBD program is translated to a compliant executable PLC code. For more details on the FBD programming language and PLCs, we refer the reader to the work of John \textit{et al.}~\cite{john2010iec}.

We experimented with 37 industrial FBD programs for which we applied the Q-EMCQ approach. These programs contain ten input parameters and 1209 lines of code on average per program.

To answer our research questions, we generated test cases using Q-EMCQ for $2-wise$, $3-wise$, and $4-wise$ and executed each program on these test cases to collect branch coverage and fault detection scores for each test suite as well as the number of test cases created. A \textit{test suite} created for a PLC program
contains a set of \textit{test cases} containing inputs, expected and actual outputs together with timing constraints.

\paragraph{Test Case Generation and Manual Testing.}
We used test suites automatically generated using Q-EMCQ. To do this, we asked an engineer from Bombardier Transportation Sweden AB, responsible for developing and testing the PLC programs used in this study, to identify the range parameter values for each input variable and constraints. We used the collected input parameter ranges for each input variable for generating combinatorial test cases using Q-EMCQ. These ranges and constraints were also used for creating manual test suites. We collected the number of test cases for each manual test suite created by engineers for each of the programs used in this case study. In testing these PLC programs, the testing processes are performed according to safety standards and certifications, including rigorous specification-based testing based on functional requirements expressed in natural language. As the programs considered in this study are manually tested and are part of a delivered project, we expect that the number of test cases created manually by experienced industrial engineers to be a realistic proxy measure of the level of efficiency needed to test these PLC programs thoroughly. 

\paragraph{Measuring Branch Coverage.}
Code coverage criteria are used in practice to assess the extent to which the PLC program has been covered by test cases \cite{ammann2008introduction}. Many criteria have been proposed in the literature, but in this study, we only focus on branch coverage criteria. For the PLC programs used in this study, the engineers developing software indicated that their certification process involves achieving high branch coverage.   
A branch coverage score was obtained for each test suite. A test suite satisfies decision coverage if running the test cases causes each branch in the program to have the value \emph{true} at least once and the value \emph{false} at least once. 

\paragraph{Measuring Fault Detection.}
Fault detection was measured using mutation analysis by generating faulty versions of the PLC programs Mutation analysis is used in our case study by creating faulty implementations of a program in an automated manner to examine the fault detection ability of a test case~\cite{demillo1978hints}. A mutated program is a new version of the original PLC program created by making a small change to this original program. For example, in a PLC program, a mutated program is created by replacing an operator with another, negating an input variable, or changing the value of a constant to another interesting value. If the execution of a test suite on the mutated program gives a different observable behavior as the original PLC program, the test case kills that mutant. We calculated the mutation score using an output-only oracle against all the created mutated programs. For all programs, we assessed the mutation detection capability of each test case by calculating the ratio of mutated programs killed to the total number of mutated programs. Researchers ~\cite{just2014mutants,andrews2005mutation} investigated the relation between real fault detection and mutant detection and there is some strong empirical evidence suggesting that if a test case can detect or kill most mutants, it can also be good at detecting naturally-occurring faults, thus providing evidence that the mutation score is a fairly good proxy measure for fault detection.

In the creation of mutants, we rely on previous studies that looked at using mutation analysis for PLC software~\cite{shin2012empirical,enoiu2017comparative}. We used the mutation operators proposed in~\cite{enoiu2017comparative} for this study. The following mutation operators were used:

\begin{itemize}
\item {\it Logic Block Replacement Operator (LRO)}. Replacing a logical block with another block from the same category (e.g., replacing an AND block with an XOR block in Figure~\ref{fig:FBDtesting}).
\item {\it Comparison Block Replacement Operator (CRO)}. Replacing a comparison block with another block from the same category (e.g., replacing a Greater-Than (GT) block with a Greater-or-Equal (GE) block in Figure~\ref{fig:FBDtesting}).
\item {\it Arithmetic Block Replacement Operator (ARO)}. Replacing an arithmetic block with another block from the same functional category (e.g., replacing a maximum (MAX) block with an addition (ADD) block).
\item {\it Negation Insertion Operator (NIO)}. Negating an input or output connection between blocks (e.g., a variable $var$ becomes NOT($var$)).
\item {\it Value Replacement Operator (VRO)}. Replacing a value of a constant variable connected to a block (e.g., replacing a constant value ($var=5$) with its boundary values (e.g., $var=6$, $var=4$)).
\item {\it Timer Block Replacement Operator (TRO)}. Replacing a timer block with another block from the same timer category (e.g., replacing a Timer-off (TOF) block with a Timer-On (TON) block in Figure~\ref{fig:FBDtesting}).
\end{itemize}

To generate mutants, each of the mutation operators was systematically applied to each program wherever possible. In total, for all of the selected programs, 1368 mutants (faulty programs based on ARO, LRO, CRO, NIO, VRO, and TRO operators) were generated by automatically introducing a single fault into the program.

\paragraph{Measuring Cost.}
Leung and White~\cite{leung1991cost} proposed the use of a cost model for comparing testing techniques by using direct and indirect testing costs. A direct cost includes the engineer's time for performing all activities related to testing but also the machine resources such as the test environment and testing tools. On the other hand, indirect cost includes test process management, tool development. To accurately measure the cost effort, one would need to measure the direct and indirect costs for performing all testing activities. However, since the case study is performed a postmortem on a system that is already in use and for which the development is finished, this type of cost measurement was not feasible. Instead, we collected the number of test cases generated by Q-EMCQ as a proxy measure for the cost of testing. We are interested in investigating the cost of using the Q-EMCQ approach in the same context as manual testing. In this case study, we consider that costs are related to the number of test cases. The higher the number of test cases, the higher is the respective test suite cost. We assume this relationship to be linear. For example, a complex program will require more effort for understanding, and also more tests than a simple program. Thus, the cost measure is related to the same factor-- the complexity of the software which will influence the number of test cases. Analyzing the cost measurement results is directly related to the number of test cases giving a picture of the same effort per created test case. In addition to the number of test cases measure, other testing costs are not considered, such as setting up the testing environment and tools, management overhead, and the cost of developing new tests. In this work, we restrict our analysis to the number of test cases created in the context of our industrial case study.

\subsubsection{Module Clustering of Class Diagrams}

The details of the three class diagrams involved are:

\begin{itemize}
\item Credit Card Payment System (CCPS) \cite{REF-1} consists of 14 classes interlink with 20 two-way associations and 1 aggregation relationship (refer to Fig.\ref{Fig9a} ).
\item Unified Inventory University ((UIU) \cite{REF-2} consists of 19 classes interlink with 28 aggregations, 1 $2-wise$ associations and 1 dependency relationship (refer to Fig. \ref{Fig10a}).
\item Food Book (FB)\footnote{https://bit.ly/2XDPOPB} consists of 31 interlinked classes with 25 $2-wise$ associations, 7 generalizations, and 6 aggregations clustered into 3 packages (refer to Fig. \ref{Fig11a}).
\end{itemize}

Module clustering problem involves partitioning a set of modules into clusters based on the concept of coupling (i.e., measuring the dependency between modules) and cohesion (i.e., measuring the internal strength of a module cluster). The higher the coupling, the less readable the piece of code will be; whereas, the higher the cohesion, the better to code organization will be. To allow its quantification, Praditwong et al. \cite{REF-4} define Modularization Quality(MQ) as the sum of the ratio of intra-edges and inter-edges in each cluster, called Modularization Factor (MFk) for cluster k based on the use of module dependency graph such as the class diagram.  Mathematically, MFk can be formally expressed as in Eq\ref{eq.11}:

\begin{equation}
\centering
    MF_{k}=\left\{\begin{matrix}
 0 & if\, i=0\\
 \frac{i}{ i+\frac{1}{2}    j    } & if\, i>0
\end{matrix}\right.
\label{eq.11}
\end{equation}

where $i$ is the weight of intra-edges and $j$ is that of inter-edges. The term $\frac{1}{2}j$ is to split the penalty of inter-edges across the two clusters that are connected by that edge. The MQ can then be calculated as the sum of $MF_k$ as follows:

\begin{equation}
\centering
    MQ= \sum _{n}^{k=1} MF_k
\end{equation}

where $n$ is the number of clusters, it should be noted that maximizing MQ, which does not necessarily mean maximizing the clusters.

\section{Case Study Results}

The case study results can be divided into two parts, namely for answering RQ1-RQ5 and for answering RQ6.

\subsection{Answering RQ1-RQ5}

This section provides an analysis of the data collected in this case study, including the efficiency of Q-EMCQ and the effectiveness of using combinatorial interaction testing of different strengths for industrial control software. For each program and each generation technique considered in this study, we collected the produced test suites (i.e., \textit{$2-wise$} stands for Q-EMCQ generated test suites using pairwise combinations, \textit{$3-wise$} is short for test suites generated using Q-EMCQ and $3-wise$ interactions and \textit{$4-wise$} stands for generated test suites using Q-EMCQ and $4-wise$ interactions). The overall results of this study are summarized in the form of boxplots in Figure~\ref{fig:AllBoxes}.  Statistical analysis was performed using the R software~\cite{anondevelopment}.

As our observations are drawn from an unknown distribution, we evaluate if there is any statistical difference between \textit{$2-wise$}, \textit{$3-wise$}, and \textit{$4-wise$} without making any assumptions on the distribution of the collected data. We use a Wilcoxon-Mann-Whitney U-test~\cite{howell2012statistical}, a non-parametric hypothesis test for determining if two populations of data samples are drawn at random from identical populations. This statistical test was used in this case study for checking if there is any statistical difference among each measurement metric. Besides, the Vargha-Delaney test~\cite{vargha2000critique} was used to calculate the standardized effect size, which is a non-parametric magnitude test that shows significance by comparing two populations of data samples and returning the probability that a random sample from one population will be larger than a randomly selected sample from the other. According to Vargha and Delaney~\cite{vargha2000critique}, statistical significance is determined when the obtained effect size is above 0,71 or below 0,29. 

For each measure, we calculated the effect size of $2-wise$, $3-wise$, and $4-wise$ and we report in Table~\ref{tab:effectsize} the p-values of these  Wilcoxon-Mann-Whitney U-tests with statistically significant effect sizes shown in bold. 

\subsection*{\textbf{RQ1: In what ways does the use of Q-EMCQ improve upon EMCQ?}}

Table~\ref{Table:MCH_and_EMCQ} highlights the results for both Q-EMCQ and EMCQ results involving the 3 combinations of mixed covering arrays $MCA (N; 2, 5^1 3^3 2^2)$, $MCA (N; 3, 5^2 4^2 3^2)$, and $MCA (N; 4, 5^1 3^2 2^3)$.

\begin{table*}

\centering

\caption{Size and Time Comparison for Q-EMCQ and its predecessor EMCQ}
\includegraphics[width= 5 in]                    {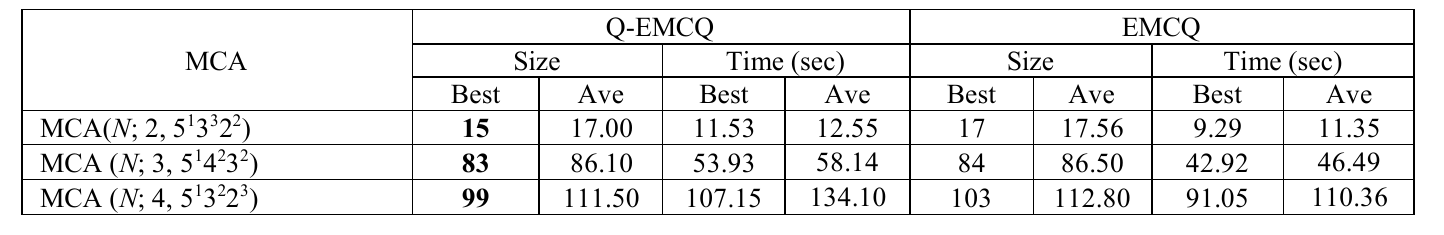}

\label{Table:MCH_and_EMCQ}
\end{table*}

Referring to Table~\ref{Table:MCH_and_EMCQ}, we observe that Q-EMCQ has outperformed EMCQ as far as the average test suite size is concerned in all three MCAs. As for the time performances, EMCQ is better than Q-EMCQ, notably because there is no overhead as far as maintaining the Q-learning table.

\begin{figure*}
	\centering
	\begin{subfigure}[b]{1\textwidth}
    \centering
		\includegraphics[width= 5 in]	{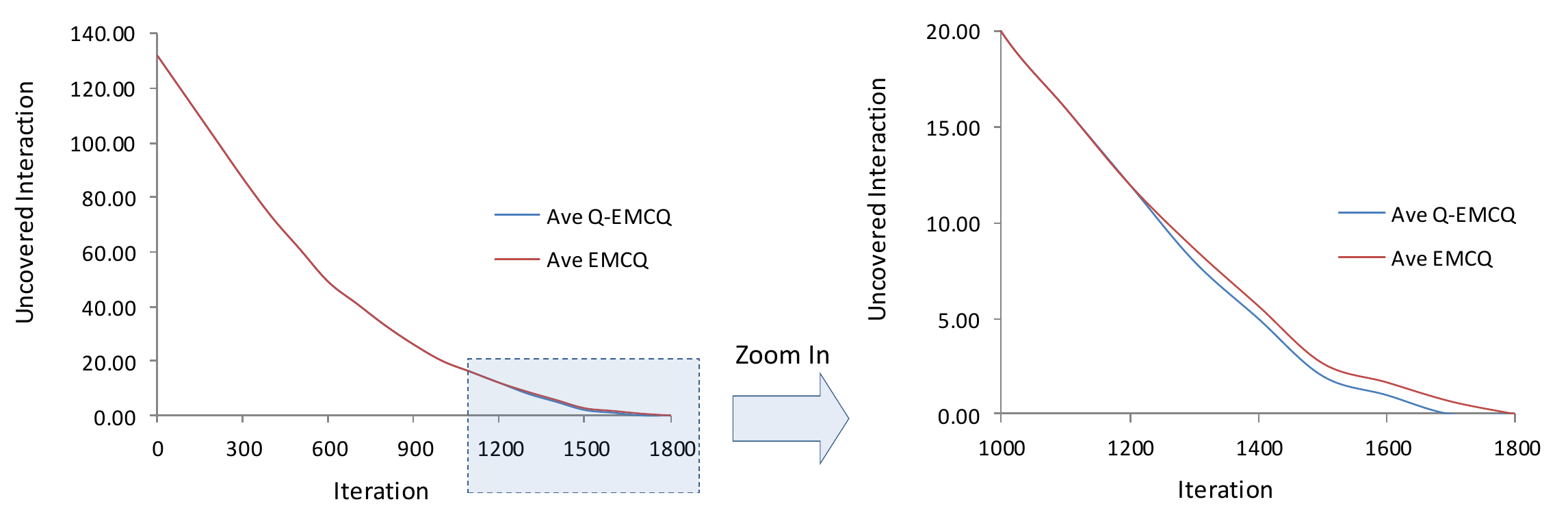}
		\caption{Average Convergence - $MCA(N; 2, 5^1 3^3 2^2)$ for Q-EMCQ and EMCQ}
		\label{Fig:Figure11}
    \end{subfigure}%
         \hfill
    \begin{subfigure}[b]{1\textwidth}
    \centering
    	\includegraphics[width= 5 in]					{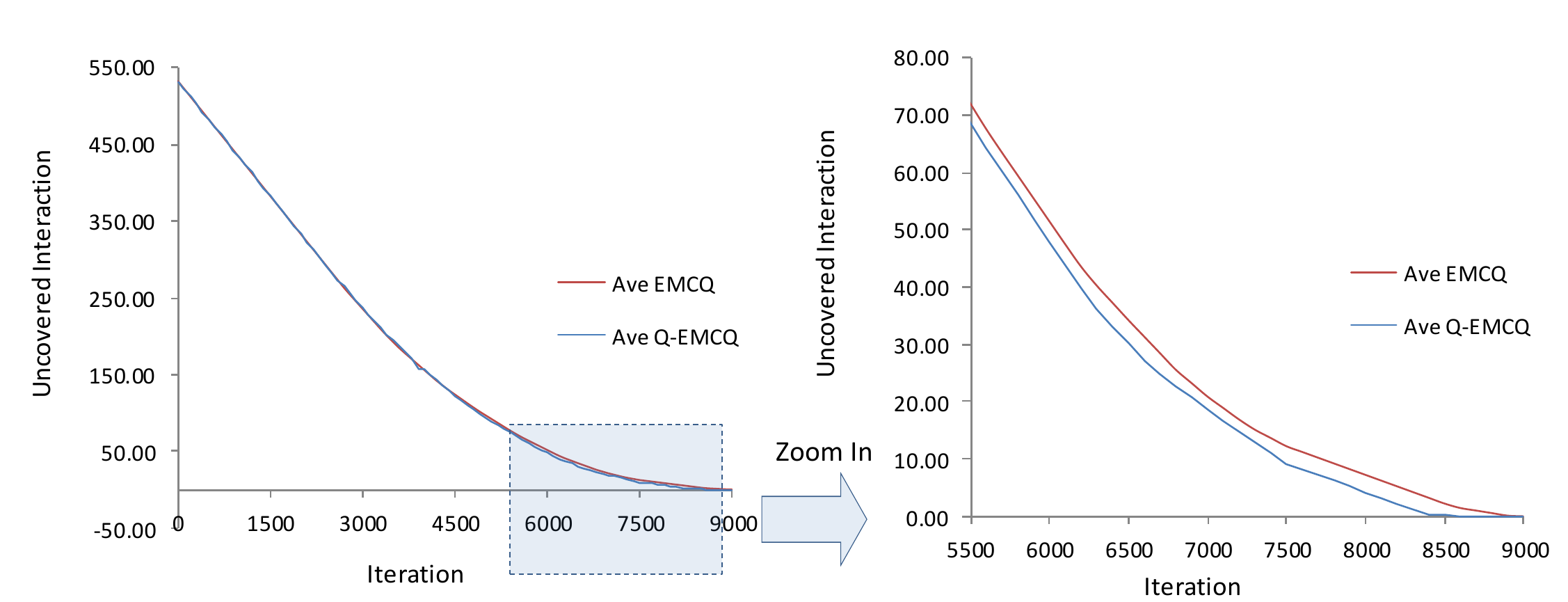}
        \caption{Average Convergence - $MCA (N; 3, 5^1 4^2 3^2)$ for Q-EMCQ and EMCQ}
        \label{Fig:Figure12}
          \end{subfigure}
 \hfill
    \begin{subfigure}[b]{1\textwidth}
    \centering
    	\includegraphics[width= 5 in]					{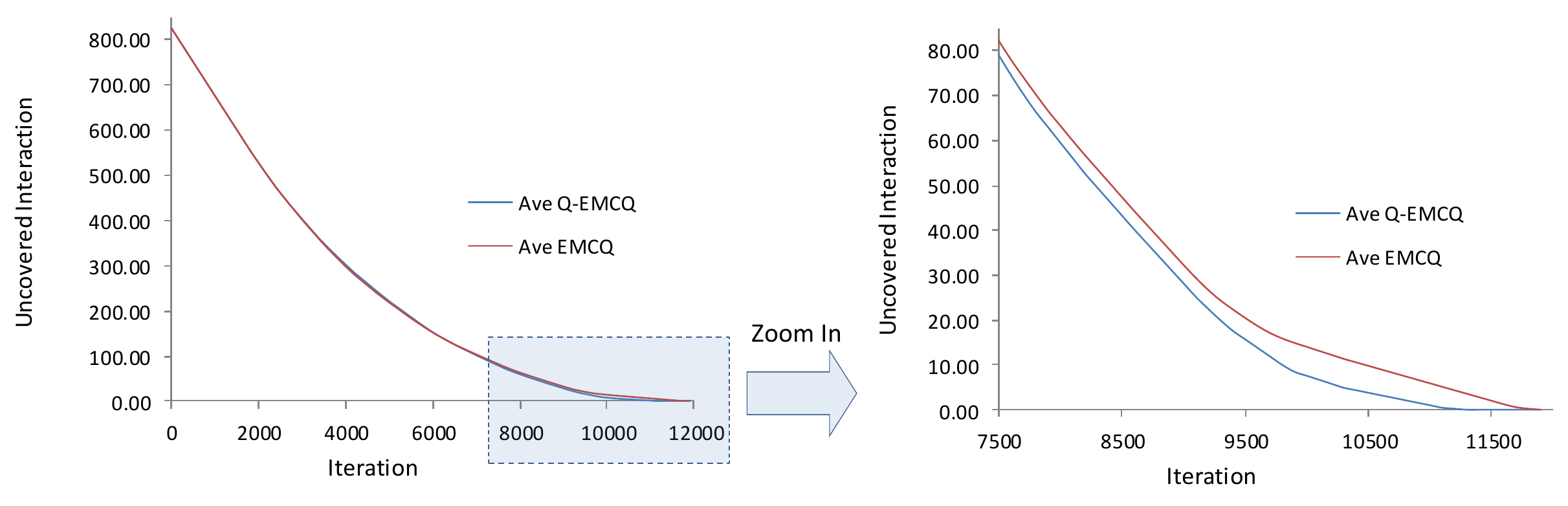}
        \caption{Average Convergence - $MCA (N; 4, 5^1 3^2 2^3)$ for Q-EMCQ and EMCQ}
        \label{Fig:Figure13}
          \end{subfigure}
\caption{Average convergences for Q-EMCQ and EMCQ for different CAs.}\label{fig:convergence}
\end{figure*}

To investigate the performance of Q-EMCQ and EMCQ further, we plot the convergence profiles for the 20 runs for the three covering arrays, as depicted in Figure~\ref{Fig:Figure11} till Figure~\ref{Fig:Figure13}. At a glance, visual inspection indicates no difference as far as average convergence is concerned. Nonetheless, when we zoom in all the figures (on the right of Figure~\ref{Fig:Figure11} till Figure~\ref{Fig:Figure13}), we notice that Q-EMCQ has better average convergence than EMCQ.

\subsection*{\textbf{RQ2: How good is the efficiency of Q-EMCQ in terms of test suite minimization when compared to existing strategies?}}

Table~\ref{Table:Table2Kamal} and~\ref{Table:Table3Kamal} highlight the results of two main experiments involving $CA (N; t, v^7)$ with variable values $2 \le v \le 5$, $t$ varied up to 4 as well as $CA (N; t, 3^k)$ with variable number of parameters $3 \le k \le 12$, $t$ varied up to 4. In general, the authors of the strategies used in our experimental comparisons only provide the best solution quality, in terms of the size N, achieved by them. Thus, these strategies cannot be statistically compared with Q-EMCQ.

\begin{table*}

\centering

\caption{$CA (N; t, v^7)$ with variable values $2 \le v \le 5$, with $t$ varied up to 4}
\includegraphics[width= 5 in]                    {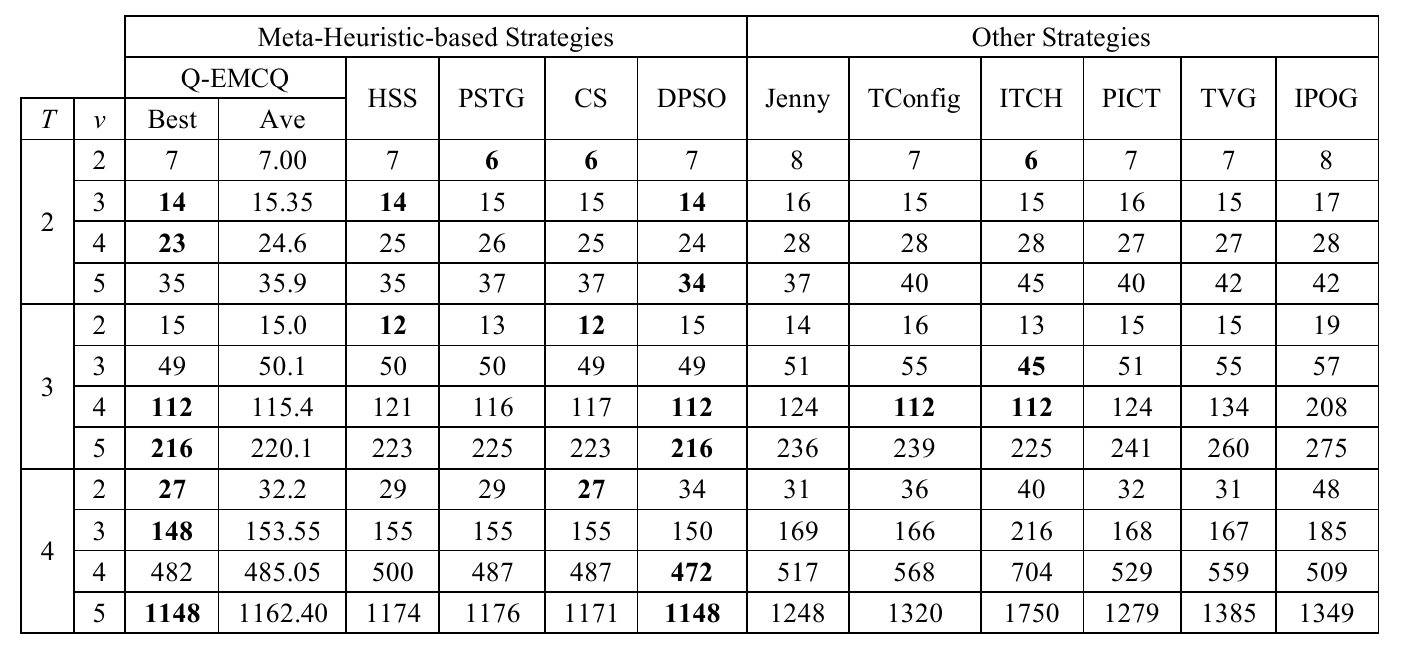}

\label{Table:Table2Kamal}
\end{table*}

\begin{table*}

\centering

\caption{$CA (N; t, 3^k)$ with variable number of parameters $3 \le k \le 12$, with $t$ varied up to 4}
\includegraphics[width= 5 in]                    {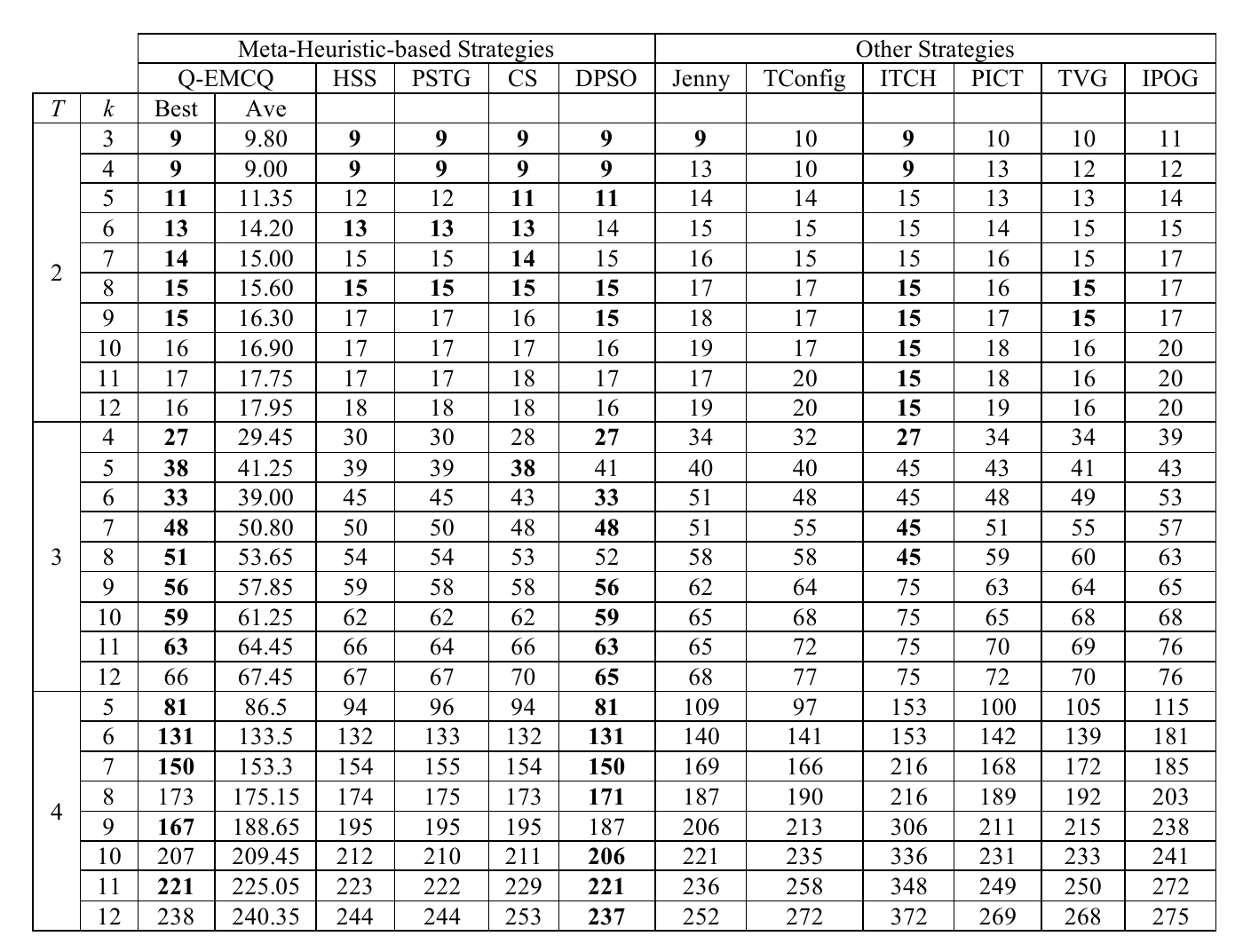}

\label{Table:Table3Kamal}
\end{table*}

As seen in Tables~\ref{Table:Table2Kamal} and~\ref{Table:Table3Kamal}, the solution quality attained by Q-EMCQ is very competitive with respect to that produced by the state-of-the-art strategies. In fact, Q-EMCQ is able to match or improve on 7 out of 16 entries in Table \ref{Table:Table2Kamal} (i.e., 43.75\%) and 20 out of 27 entries in Table \ref{Table:Table3Kamal} (i.e., 74.07\%)  respectively. The closest competitor is that of DPSO which scores 6 out of 16 entries in Table \ref{Table:Table2Kamal} (i.e., 37.50\%) and 19 out of 27 entries in Table \ref{Table:Table3Kamal} (i.e., 70.37\%). Regarding the computational effort, as the strategies used in our comparisons adopt different running environments, data structures, and implementation languages, these algorithms cannot be directly compared with ours.

\subsection*{\textbf{RQ3: How good are combinatorial tests created using Q-EMCQ for 2-wise, 3-wise and 4-wise at covering the code?}}

In Table~\ref{tab:resultsAll} we present the mutation scores, code coverage results and the number of test cases in each collected test suite (i.e., $2-wise$, $3-wise$ and $4-wise$ generated tests). This table lists the minimum, maximum, median, mean, and standard deviation values. To give an example, $2-wise$ created test suites found an average mutation score of 52\%, while $4-wise$ tests achieved an average mutation score of 60\%. This shows a considerable improvement in the fault-finding capability obtained by $4-wise$ test suites over their $2-wise$ counterparts. For branch coverage, combinatorial test suites are not able to reach or come close to achieving 100\% code coverage on most of the programs considered in this case study. 

\begin{table*}
\renewcommand{\arraystretch}{1.0}
\caption{Results for each measure: mutation score, branch coverage score and the cost in terms of the number of test cases. We report the cost comparison between manual tests and t-wise ($t\leq4$) generated tests. We report several statistics relevant to the obtained results: minimum, maximum, median, mean and standard deviation (SD) values.}
\label{tab:resultsAll}
\centering
\includegraphics[width=4.2 in]                {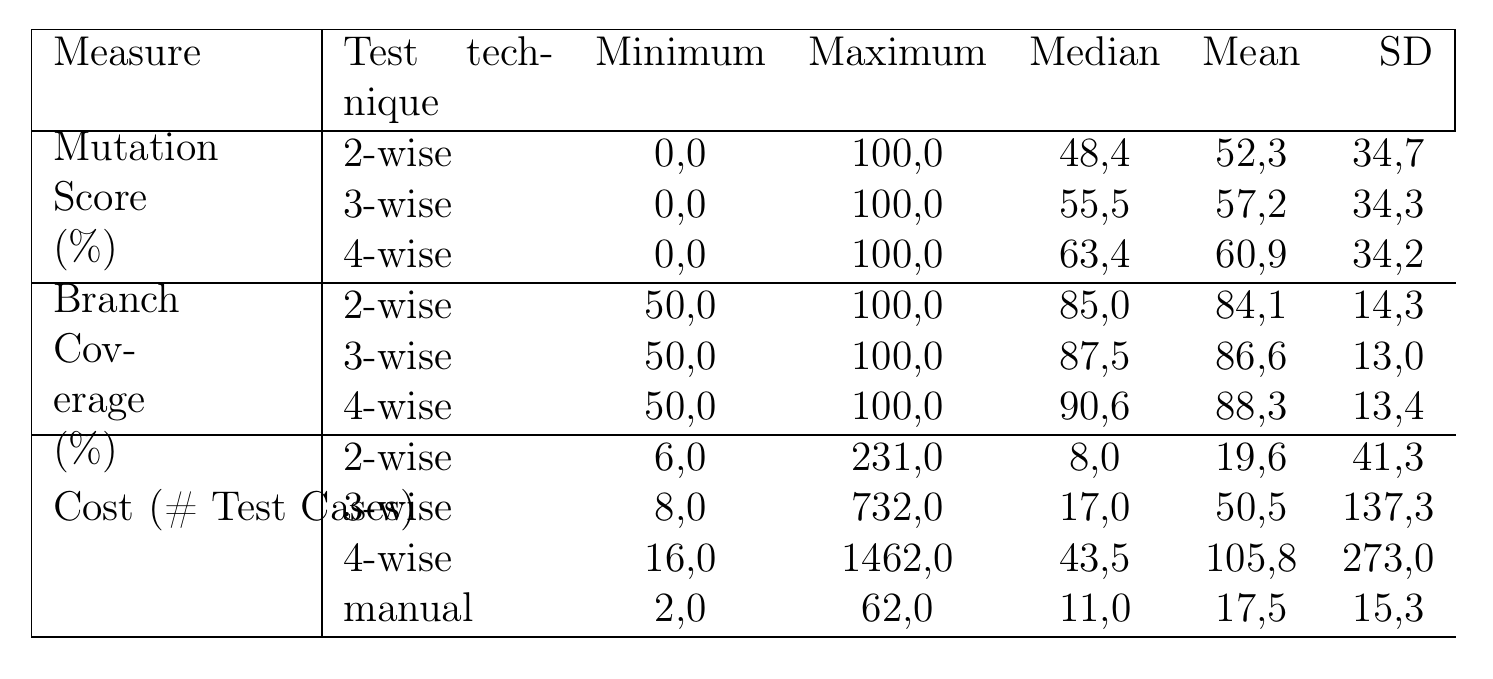}
\end{table*}

\begin{table*}
\renewcommand{\arraystretch}{1.0}
\caption{For mutation score and coverage we calculated the effect size of $2-wise$ versus $3-wise$ and $4-wise$. In addition, for the cost measure we calculated the effect size between manual testing and $t-wise$ with $t\leq4$. We also report the p-values of a Wilcoxon-Mann-Whitney U-tests with significant effect sizes shown in bold.}
\centering
\begin{tabular}{|p{0.20\linewidth}| p{0.2\linewidth} p{0.2\linewidth} p{0.2\linewidth}|} \hline
Measure & Method & Effect Size & \hspace{0,03cm} p-value\\
    \hline
\multirow{6}{*}{\parbox{0.5cm}{Mutation Score}} & 2-wise & \hspace{0,25cm} 0.444 & \hspace{0,25cm} 0.411\\ 

&3-wise & &\\ \cline{2-4}
    & 3-wise &  \hspace{0,25cm} 0.464 & \hspace{0,25cm} 0.595 \\ 
&  4-wise&  &  \\ \cline{2-4}
& 2-wise &  \hspace{0,25cm} 0.412 & \hspace{0,25cm} 0.193 \\ 
&  4-wise&  &  \\ \cline{2-4}
  \hline 
 \multirow{6}{*}{\parbox{1.5cm}{Coverage}} & 2-wise & \hspace{0,25cm} 0.454 & \hspace{0,25cm} 0.494\\ 
 & 3-wise & &\\ \cline{2-4}
   & 3-wise &  \hspace{0,25cm} 0.447 & \hspace{0,25cm} 0.432 \\ 
&  4-wise&  &  \\ \cline{2-4}
   & 2-wise &  \hspace{0,25cm} 0.412 & \hspace{0,25cm} 0.186 \\ 
&  4-wise&  &  \\ \cline{2-4}
    \hline 
     \multirow{6}{*}{Cost (\#test cases)} 
    &  manual & \hspace{0,25cm} 0.536 &  $<$ 0,565\\ 
     
& 2-wise & &\\ \cline{2-4}
& manual &  \hspace{0,25cm} 0,376 & $<$ 0,054 \\ 
&  3-wise&  &  \\ \cline{2-4}
   & manual &  \hspace{0,25cm} \textbf{0.157} & $<$ 0.001 \\ 
&  4-wise&  &  \\ \cline{2-4}
    \hline 
\end{tabular}
    \label{tab:effectsize}
\end{table*}

\begin{figure*}
    \centering
    \begin{subfigure}[b]{0.40\textwidth}
        \includegraphics[width=\textwidth]                    {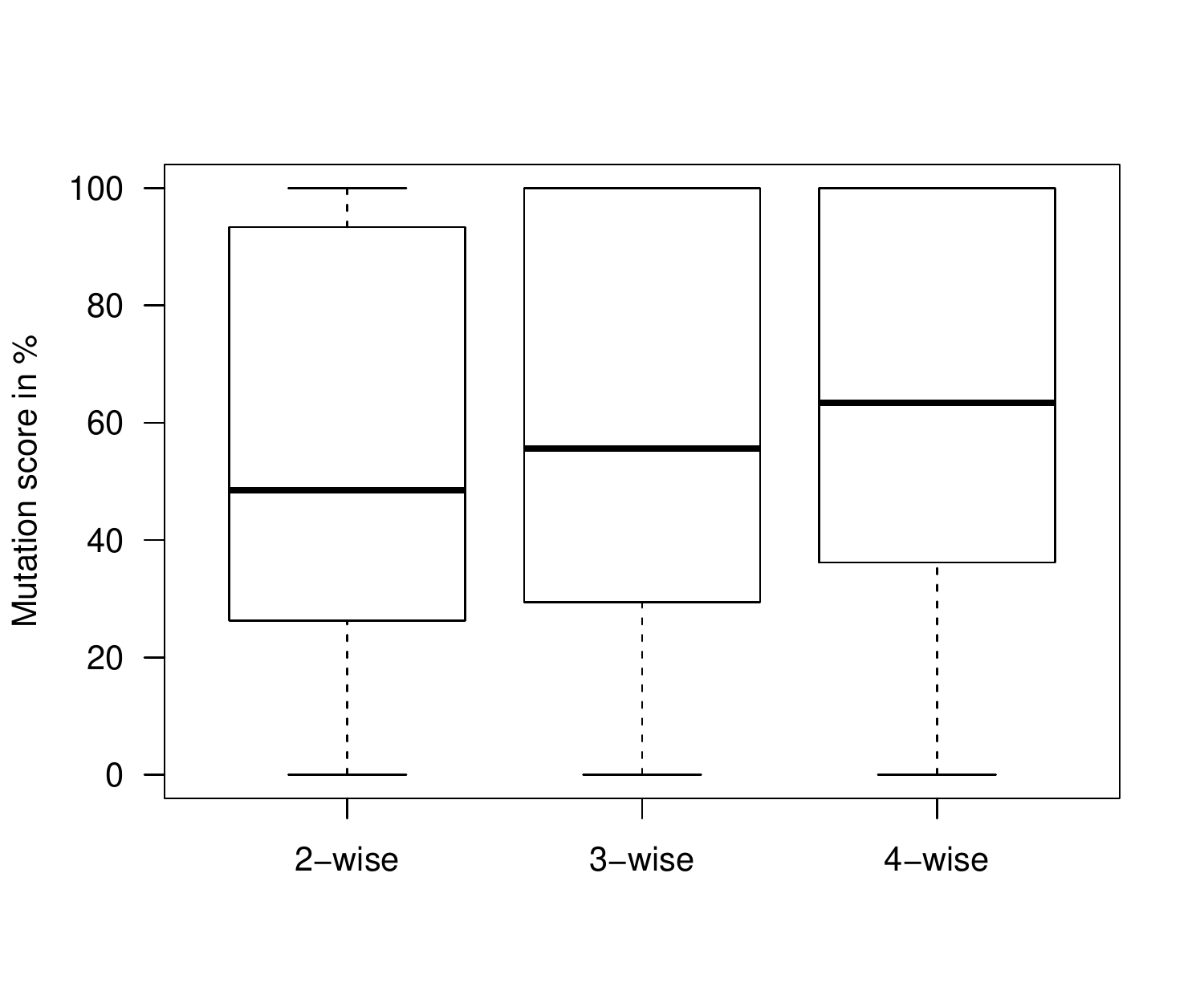}
                \vspace{-1.10cm}
        \caption{Mutation Score}
        \label{mutationscoreRand}
    \end{subfigure}%
         \hfill
    \begin{subfigure}[b]{0.40\textwidth}
        \includegraphics[width=\textwidth]                    {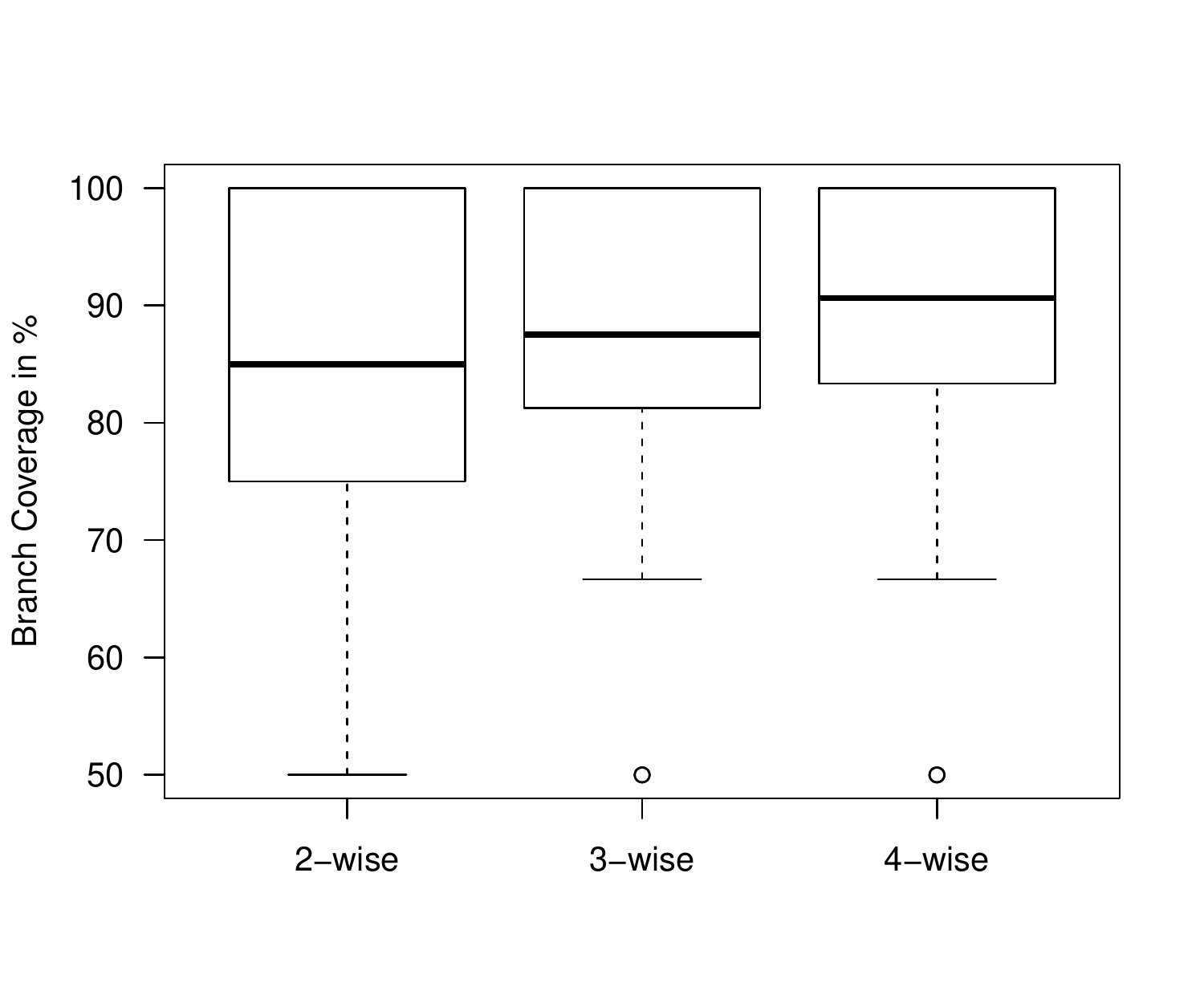}
                \vspace{-1.10cm}
        \caption{Branch Coverage}
        \label{fig:decisioncoverageRand}
          \end{subfigure}
\caption{Mutation score and achieved branch coverage comparison between $2-wise$, $3-wise$ and $4-wise$ generated test cases; boxes span from 1st to 3rd quartile, black middle lines mark the median, and the whiskers extend up to 1.5x the interquartile range and the circle symbols represent outliers.}\label{fig:AllBoxes}
\end{figure*}

As seen in Figure~\ref{fig:decisioncoverageRand}, for the majority of programs considered, combinatorial test suites achieve at least 50\% branch coverage. $2-wise$ test suites achieve lower branch coverage scores (on average 84\%) than $3-wise$ test suites (on average 86\%). The coverage achieved by combinatorial test suites using $4-wise$ is ranging between 50\% and 100\% with a median branch coverage value of 90\%. 

As seen in Figure~\ref{fig:decisioncoverageRand}, the use of combinatorial testing achieves between 84\% and 88\% branch coverage on average. Results for all programs (in Table~\ref{tab:effectsize}) show that differences in code coverage achieved by $2-wise$ versus $3-wise$ and $4-wise$ test suites are not strong in terms of any significant statistical difference (with an effect size of 0.4). Even if automatically generated test suites are created by having the purpose of covering up to $4-wise$ input combinations, these test suites are not missing some of the branches in the code. The results are matching our expectations: combinatorial test suites achieve high code coverage to automatically generated test suites using combinatorial goals up to $4-wise$ achieve high branch coverage. Nevertheless, we confirm that there is a need to consider other test design aspects and higher $t-wise$ strengths to achieve over 90\% branch coverage. This underscores the need to study further how combinatorial testing can be improved in practice and what aspects can be taken into account to achieve better code coverage. The programs considered in this study are used in real-time systems to provide operational control in trains. The runtime behavior of such systems depends not only on the choice of parameters but also on providing the right choice of values at the right time-points. By consider such information, combinatorial tests might be more effective at covering the code. This needs to be further studied by considering the extent to which $t-wis$e can be used in combination with other types of information.

\begin{figure}

    \centering
        \includegraphics[width=0.35\textwidth]                    {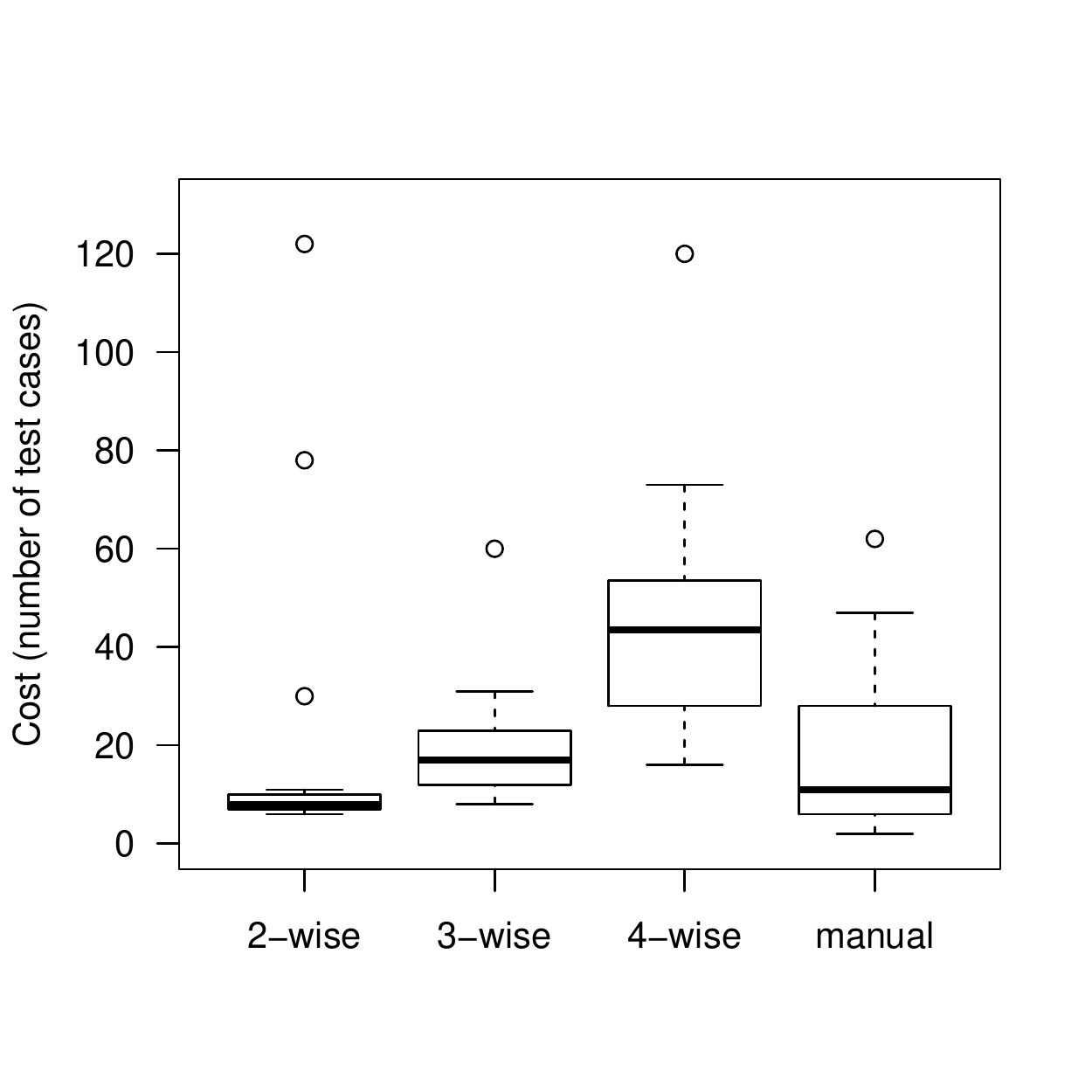}

\caption{Cost comparison in terms of number of test cases between $2-wise$, $3-wise$, $4-wise$ generated test cases and manual testing testing.}\label{fig:numberofTests}
\end{figure}

\subsection*{\textbf{RQ4: How effective are tests generated using Q-EMCQ for 2-wise, 3-wise, and 4-wise at detecting injected faults?}}
To answer RQ4 regarding the effectiveness in terms of fault detection, we focused on analyzing the test suite quality of combinatorial testing. For all programs, as shown in Figure~\ref{mutationscoreRand}, the fault detection scores of pairwise generated test suites are showing an average mutation score of 52\% but they are not significantly worse than $3-wise$ (57\% on average) and $4-wise$ (60\%) test suites with no statistically significant differences (effect size of 0,4 in Table~\ref{tab:effectsize}). Hence, a test is generated automatically using combinatorial techniques up to $4-wise$ is not a good indicator of test effectiveness in terms of mutation score. However, one hypothesis is emerging from this result: If $4-wise$ test suites are not achieving a high mutation score, there is a need to generate higher strength test suites as well as find ways to improve the fault detection scores by using other test design techniques. 

This is, to some extent, an entirely surprising result. Our expectation was that combinatorial testing of higher strength than $2-wise$ would yield high scores (over 90\%) in terms of fault detection. Tests for $4-wise$ in testing FBD programs would intuitively be quite good test cases at detecting faults. However, the results of our study are not consistent with the results of other studies~\cite{kuhn2010sp,kuhn2004software,kuhn2002investigation} reporting the degree of interaction occurring in naturally-occurring faults. Surprisingly, this expectation does not clearly hold for the results of this study. Our results indicate that combinatorial test cases with interactions up to $4-wise$ are not good indicators of test effectiveness in terms of fault detection. In addition, our results are not showing any statistically significant difference in mutation score between any $t-wise$ strength considered in this study.

\subsection*{\textbf{RQ5: How do Q-EMCQ for $2-wise$, $3-wise$ and $4-wise$ compare with manual testing in terms of cost?}}
As a baseline for comparing the cost of testing, we used test cases created by industrial engineers in Bombardier Transportation for all 37 programs included in this case study. These programs are part of a project delivered already to customers and thoroughly tested. Each test suite contains a set of test cases containing inputs, expected and actual outputs, and time information expressing timing constraints. As in this case study we consider the number of test cases related to the cost of creating, executing and checking the result of each test case, we use the number of test cases in a test suite manually created as a realistic measure of cost encountered in the industrial practice for the programs considered. We assume that the higher the number of test cases, the higher are the respective cost associated with each test suite. This section aims to answer RQ5 regarding the relative cost of performing testing concerning the number of test cases generated using Q-EMCQ in comparison with manually hand-crafted tests. As seen in Table~\ref{tab:resultsAll}, the number of test cases for $2-wise$ and $3-wise$ is consistently significantly lower than for $4-wise$ created tests. As seen in Table~\ref{tab:effectsize}, the cost of performing testing using Q-EMCQ for $4-wise$ is consistently significantly higher (in terms of the number of test cases) than for manually created test suites; $3-wise$ and $4-wise$ generated test suites are longer (88 and 33 more test cases on average respectively) over manual testing. There is enough evidence to claim that the results between $4-wise$ and manual test suites are statistically significant, with a p-value below the traditional statistical significance limit of 0,05 and a standardized effect size of 0,157. The effect is weaker for the result between $3-wise$ and manual test suites with a p-value of 0,05 and an effect size of 0,376.

As seen in Figure~\ref{fig:numberofTests}, the use of $2-wise$ consistently results in shorter test suites for all programs than for $3-wise$ and $4-wise$. It seems like $2-wise$ test suites are comparable with manual test suites in terms of the number of test cases. Examining Table~\ref{tab:effectsize}, we see the same pattern in the statistical analysis: standardized effect sizes being higher than 0,1, with p-value higher than the traditional statistical significance limit of 0,05. The effect is the strongest for the $2-wise$ and $4-wise$ with a standardized effect size of 0,08. It seems that $4-wise$ will create much more tests than $2-wise$, which in practice can affect the cost of performing testing.

\subsection{Answering RQ6}

As highlighted earlier, the experiment for RQ6  investigates the performance of Q-EMCQ against some selected meta/hyper-heuristics.

\subsection*{\textbf{RQ6: Apart from the minimization problem (i.e., $t-wise$ test generation), is Q-EMCQ sufficiently general to solve (maximization) optimization problem (i.e., module clustering)?}}

As a general observation from the results shown in Table \ref{Table6}, we note that hyper-heuristics generally outperform meta-heuristics. This could be due to the fact hyper-heuristics can adaptively choose the right operator based on the need for the current search. However, in terms of execution times, general meta-heuristics appear to be slightly faster than their hyper-heuristic counter-parts owing to the direct link from the problem domain to the actual search operators.

\begin{figure*}
	\centering
	\begin{subfigure}[b]{0.50\textwidth}
		\includegraphics[width=\textwidth]					{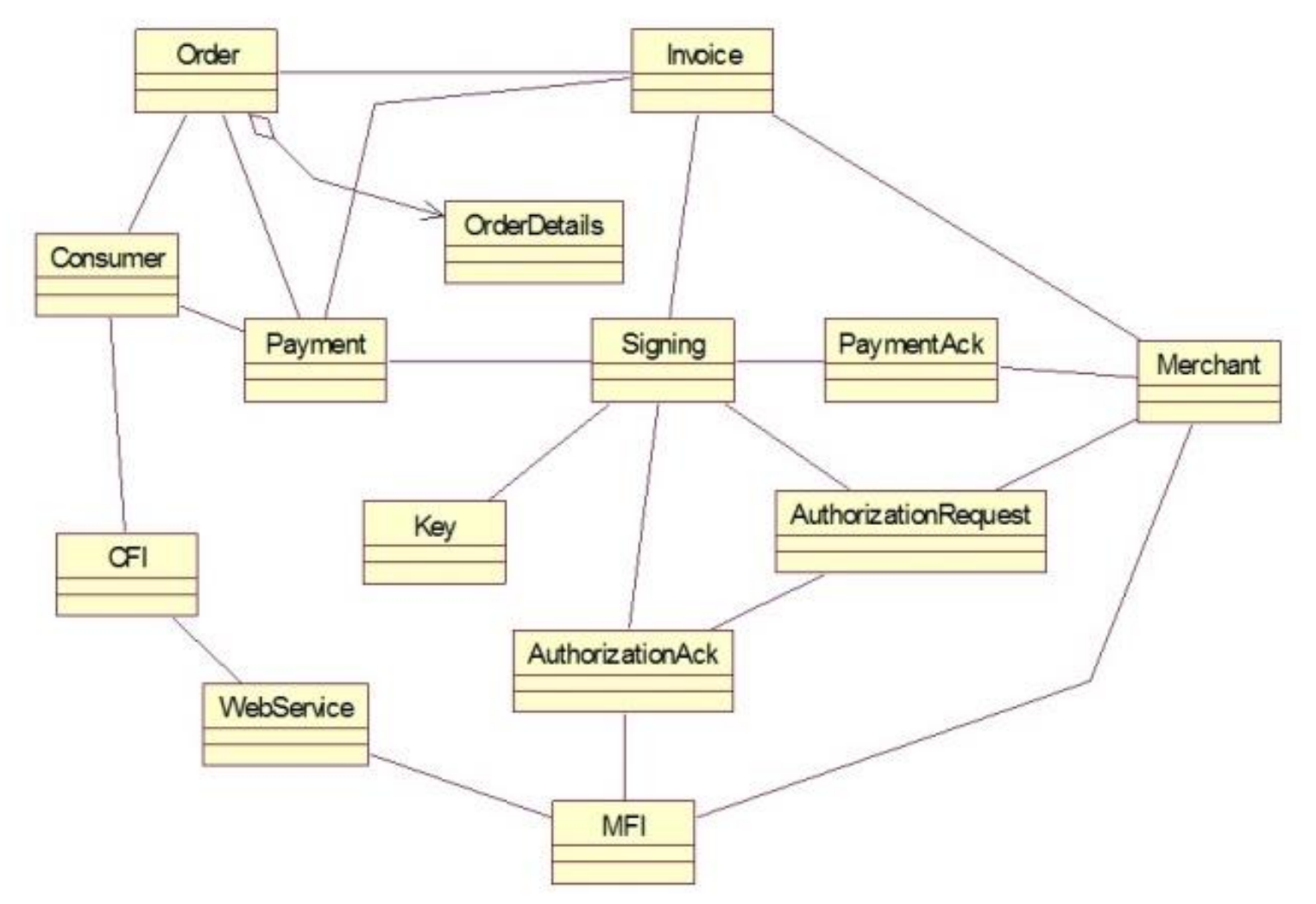}
                \vspace{-0.3cm}
		\caption{Credit Card Payment System Class Diagram}
		\label{Fig9a}
    \end{subfigure}%
         \hfill
    \begin{subfigure}[b]{0.50\textwidth}
    	\includegraphics[width=\textwidth]					{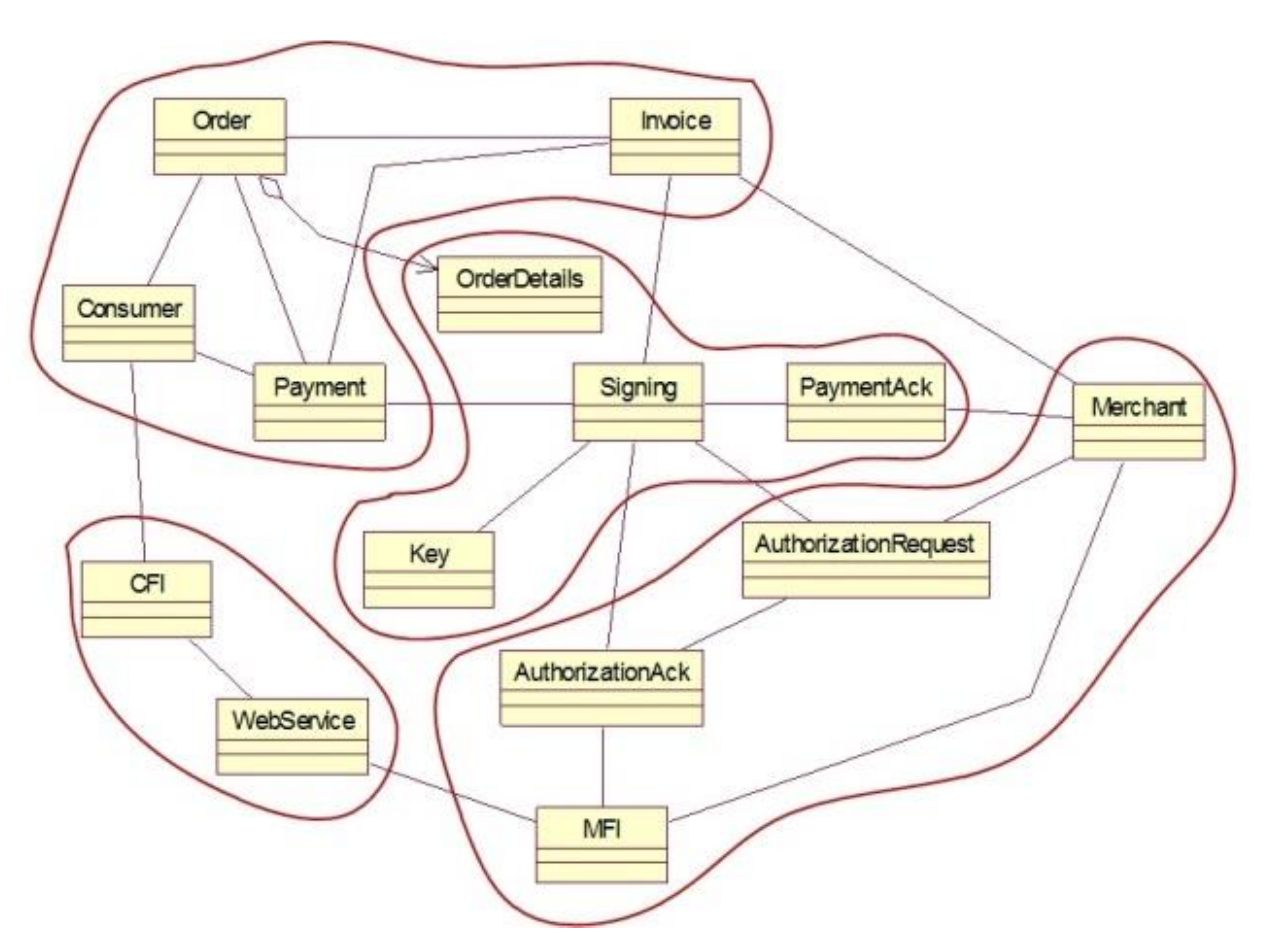}
                \vspace{-0.3cm}
        \caption{Clustering of Credit Card Payment System Class Diagram with 4 clusters and Best MQ = 2.226 generated by Q-EMCQ}
        \label{Fig9b}
          \end{subfigure}
\caption{Clustering of Case Study 1- Credit Card Payment System }\label{Figure9}
\end{figure*}

\begin{figure*}
	\centering
	\begin{subfigure}[b]{0.50\textwidth}
		\includegraphics[width=\textwidth]					{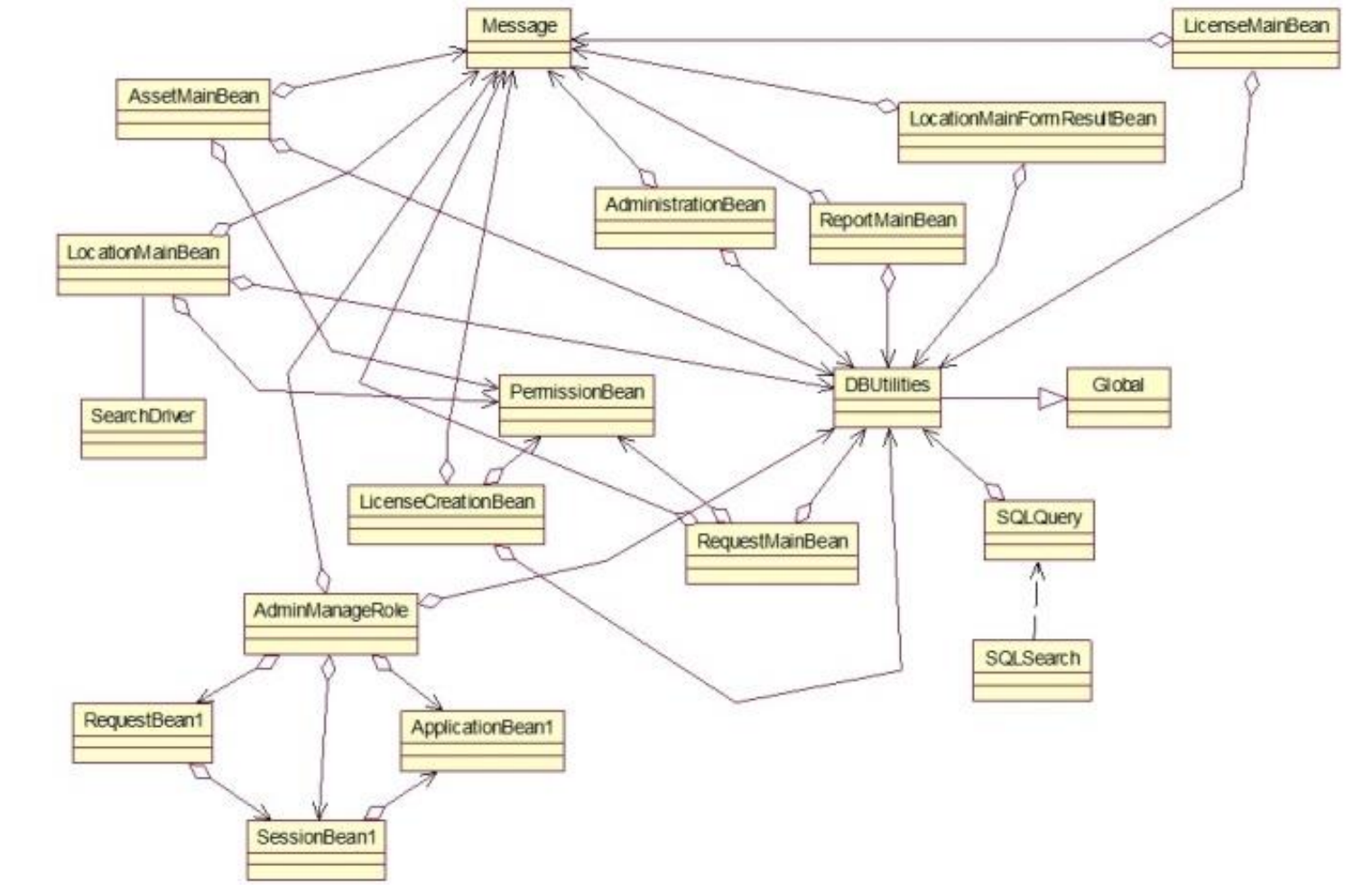}
                \vspace{-0.3cm}
		\caption{Unified University Inventory Class Diagram}
		\label{Fig10a}
    \end{subfigure}%
         \hfill
    \begin{subfigure}[b]{0.50\textwidth}
    	\includegraphics[width=\textwidth]					{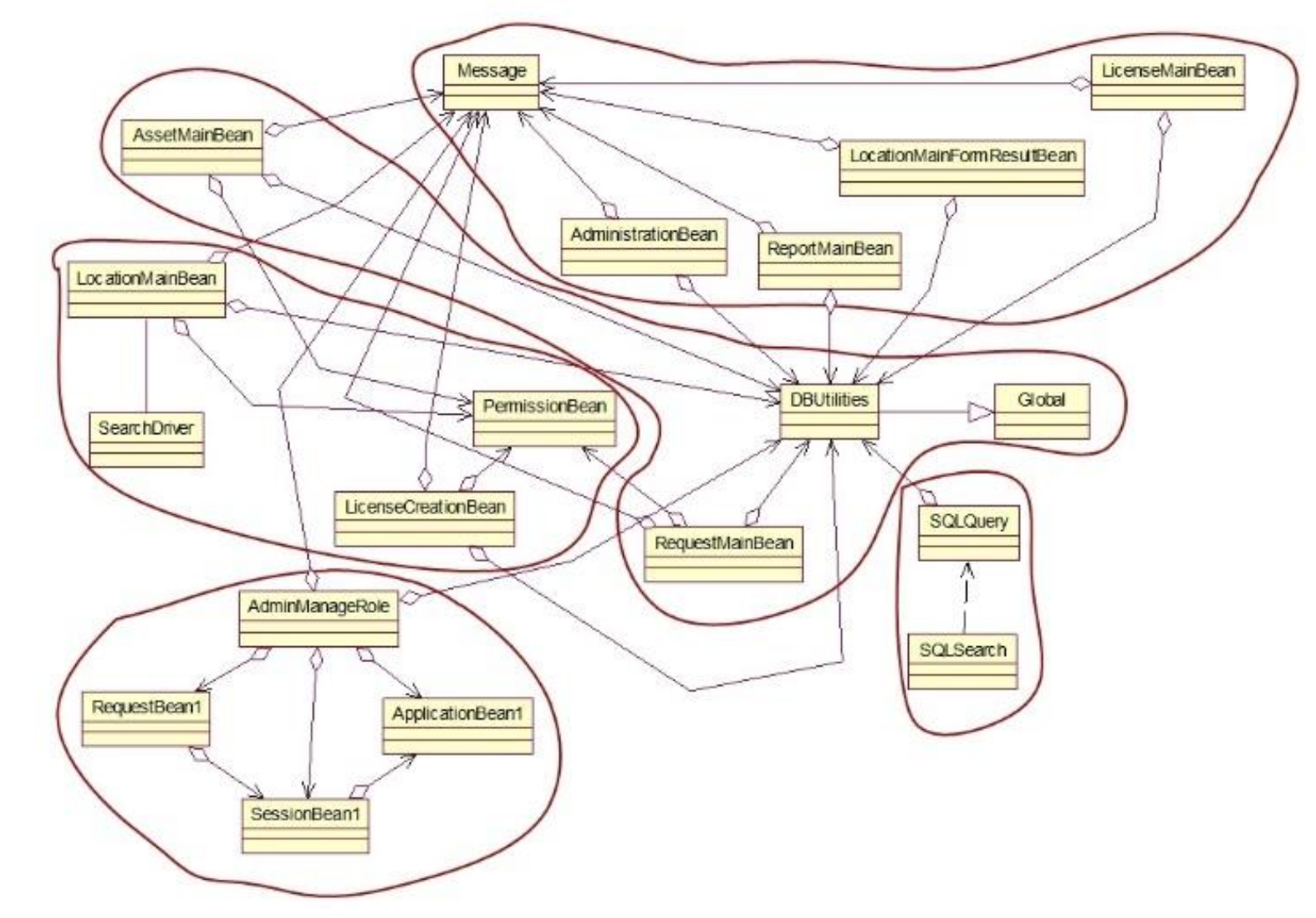}
                \vspace{-0.3cm}
        \caption{Unified University Inventory Class Diagram with 5 clusters and Best MQ = 2.899 generated by Q-EMCQ}
        \label{Fig10b}
          \end{subfigure}
\caption{Clustering of Case Study 2- Unified University Inventory}\label{Figure10}
\end{figure*}

\begin{figure*}
	\centering
	\begin{subfigure}[b]{0.48\textwidth}
		\includegraphics[width=\textwidth]					{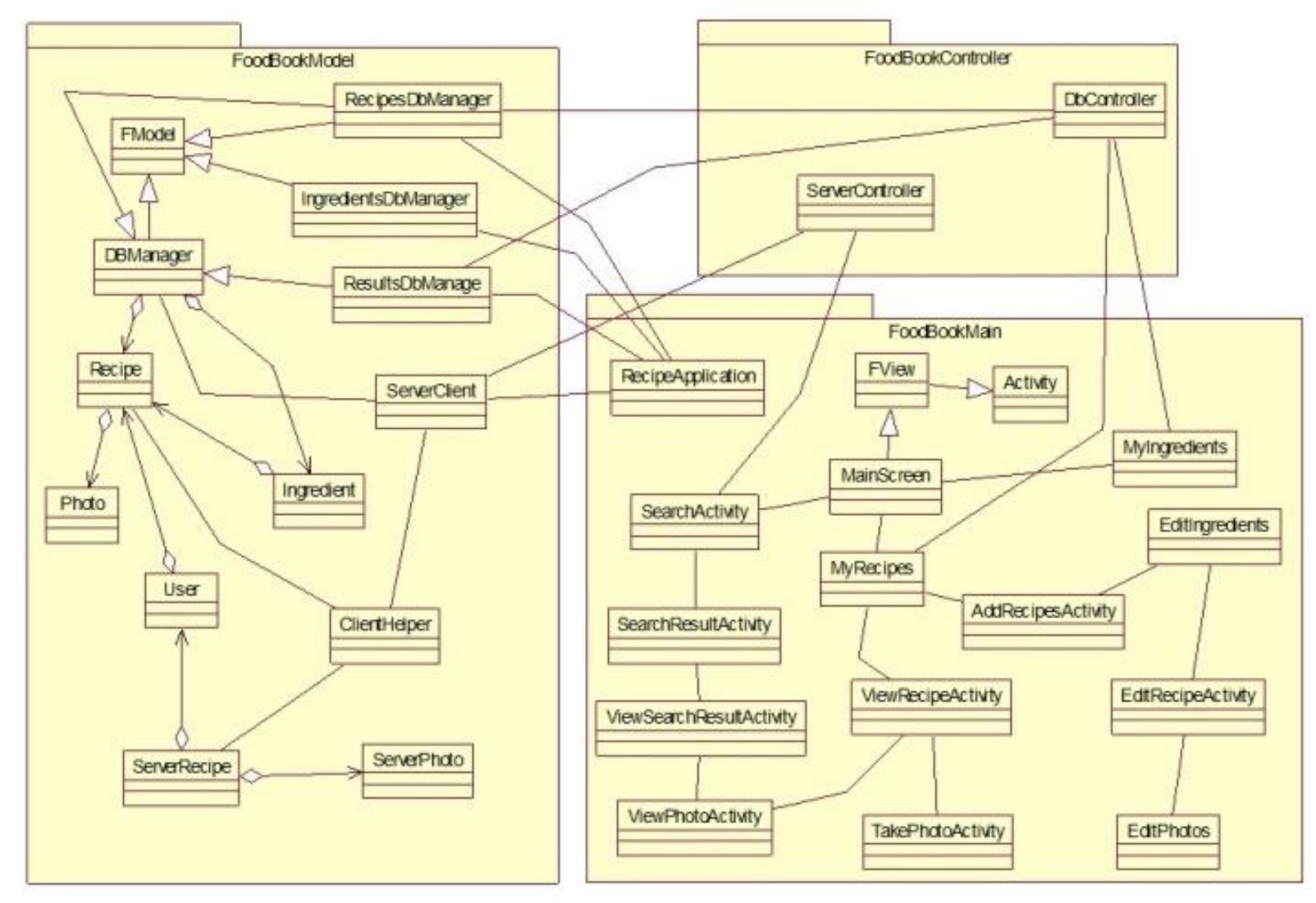}
                \vspace{-0.3cm}
		\caption{Food Book Class Diagram Programmer’s Package with 3 clusters and MQ = 1.524}
		\label{Fig11a}
    \end{subfigure}%
         \hfill
    \begin{subfigure}[b]{0.48\textwidth}
    	\includegraphics[width=\textwidth]					{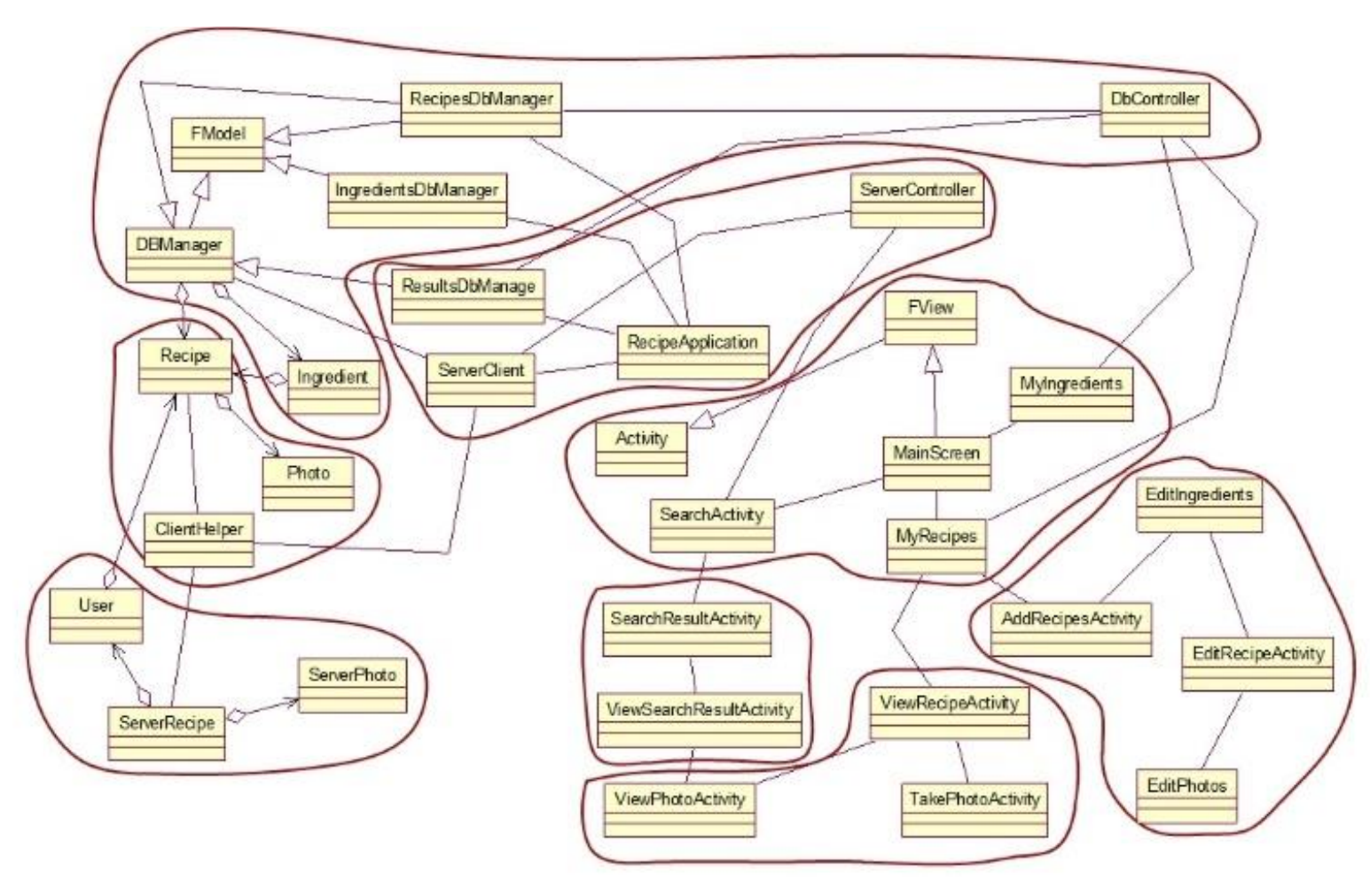}
                \vspace{-0.3cm}
        \caption{Food Book Class Diagram with 8 clusters and Best MQ = 4.466 generated by Q-EMCQ}
        \label{Fig11b}
          \end{subfigure}
\caption{Clustering of Case Study 3- Food Book Class Diagram}\label{Figure11}
\end{figure*}

Regarding the specific comparison of the hyper-heuristic group in Table \ref{Table6} and Figures \ref{Fig9b}, \ref{Fig10b} and \ref{Fig11b},  Q-EMCQ and MCF outperform all other hyper-heuristics as far as the best MQ (with 2.226, 2.899, and 4.465) for the Credit Card Payment System, Unified University Inventory and Food Book respectively. On average, Q-EMCQ has a better performance than that of MCF. Putting Q-EMCQ and MCF aside, Tabu HHH outperforms EMCQ in both average and best MQ. On the positive note, EMCQ outperforms all other hyper-heuristics as far as execution times are concerned.

Considering the comparison with the meta-heuristics, Q-EMCQ still manages to outperform all algorithms. In the case of the Credit Card Payment System, TLBO manages to match the best of MQ for Q-EMCQ, although with poorer average MQ. This is expected as the Credit Card Payment System consists of only 14 classes as compared to 19 and 31 classes in the Unified Inventory System and Food Book, respectively. In terms of execution time, SCA has the best time performance overall for Unified Inventory University (with 37.782 secs) and Food Book (with 56.798 secs) while TLBO gives the best performance for Credit Card Payment System (with 33.531 secs). Here, SOS gives the poorest execution time.

\begin{table*}

\centering

\caption{Comparing Q-EMCQ with contemporary meta/hyper-heuristics, EMCQ[52, 9], Modified Choice Function \cite{Drake2015}, TABU HHH\cite{Zamli:2016:TSH}, TLBO\cite{Rao:2011:TON}, SCA\cite{REF-5}, SOS \cite{REF-6}}.
\includegraphics[width= 6 in]                    {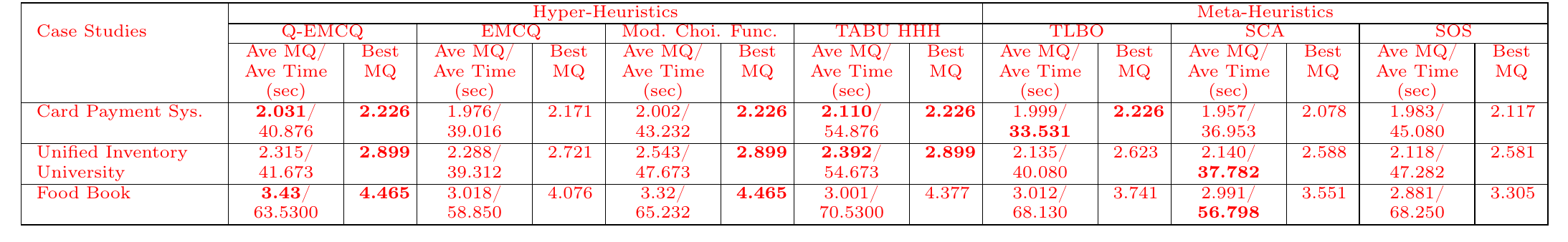}

\label{Table6}
\end{table*}

\section{Discussion} \label{Discussion}

Reflecting on the work undertaken, certain observations can be elaborated as lessons learned. In particular, we can group our observations into two parts. The first part relates to the design of Q-EMCQ and its operators, whereas the second part relates to its performance in the industrial case study.

Concerning the first part, we foresee Q-EMCQ as a general hybrid meta-heuristic. Conventional hybrid meta-heuristics are often tightly coupled (whereby two or more operators are interleaved together) and too specific for a particular problem. In addition, the selection of a particular operator during the searching process does not consider the previous performances of that operator. Contrary to conventional hybrid meta-heuristic, apart from being adaptive, Q-EMCQ design is highly flexible. Two aspects of Q-EMCQ can be treated as “pluggable” components. First, the current Monte Carlo heuristic selection and acceptance mechanism can be replaced with other selection and acceptance mechanisms. Second, the individual search operators can also be replaced with other operators (taking into consideration whether it is for local or global search). For instance, the cuckoo’s perturbation operator can easily be substituted by the simulated annealing’s neighborhood search operator. 

Unlike pure meta-heuristic approaches, Q-EMCQ also does not require any specific tuning apart from calibrating maximum iteration and population size. Notably, cuckoo as a standalone algorithm requires the calibration of three control parameters; maximum iteration, population size, and probability ($p_a$) for replacing poor eggs. Similarly, Flower as a standalone algorithm requires the calibration of three control parameters; maximum iteration, population size, and probability ($p$) for local or global pollination. Unlike the cuckoo and flower algorithms, the Jaya algorithm does not require additional parameters (other than maximum iteration and population size). Adopted as individual search operators, the cuckoo’s probability ($p_a$) and the flower’s probability ($p$) are completely abandoned within the design of Q-EMCQ. 

Similar to its predecessor EMCQ, the selection of the search operators at any instance of the searching process is adaptively performed based on the Monte Carlo heuristic selection and acceptance mechanism. However, unlike EMCQ, Q-EMCQ also keeps the memory of the best performing operators via the Q-learning table. The effect of maintaining the memory can be seen as far as average convergence is concerned. In the early iteration stage, Q-EMCQ behaves like EMCQ as far average convergence is concerned. However, toward the end of the iteration stage, while EMCQ relies solely on the random selection of operators, Q-EMCQ uses historical performance to perform the selection. For this reason, Q-EMCQ has better average convergence than EMCQ.

As far as comparative benchmark experiments with other strategies are concerned, we note that Q-EMCQ and DPSO give the best results overall (see Tables 2 and 3). On the negative note, the approach taken by DPSO is rather problem-specific. On the contrary, our experiments with maximization problems (e.g., module clustering) indicate that the Q-EMCQ approach is sufficiently general (refer to Table \ref{Table6}) although with small-time penalty to maintain the Q-learning mechanism. Here, DPSO has introduced two new control parameters as probabilities (pro1 and pro2) in addition to the existing social parameters ($c_1$ and $c_2$) and inertia weight (w) to balance between exploration and exploitation in the context of its application for $t-wise$ test generation.  In this manner, adopting DPSO to other optimization problems can be difficult owing to the need to calibrate and tune all these control parameters accordingly.

On the other side of the spectrum, PICT and IPOG appear to perform the poorest (with no results matching any of the best sizes). A more subtle observation is the fact that meta-heuristic and hyper-heuristic based strategies appear to outperform general computational based strategies.

As part of our study, we used the number of test cases to estimate the cost in terms of creation, execution, and result checking. While the cost of creating and executing a test for creating combinatorial tests can be low compared to manual testing, the cost of evaluating the test result is usually human-intensive. Our study suggests that combinatorial test suites for $4-wise$ contain 100 created test steps (number of tests) on average. By considering generating optimized or shorter test suites, one could improve the cost of performing combinatorial testing. We note here that the cost of testing is heavily influenced by the human cost of checking the test result. In this paper, we do not take into account the time of checking the results per test case. In practice, this might not be the real situation. A test strategy, which requires every input parameter in the program to be used in a certain combination, could contain test cases that are not specified in requirements. This might increase the cost of checking the test case result. A more accurate cost model would be needed to obtain more confidence in the results.

The results of this paper show that 2 to $4-wise$ combinations of values are not able to detect more than 60\% of injected faults (52\% on average for $2-wise$, 57\% on average for 3-wise, and 60\% on average for $4-wise$) and are not able to cover more than 88\% of the code (84\% on average for $2-wise$, 86\% on average for $3-2-wise$, and 88\% on average for $4-wise$). Surprisingly, these results are not consistent with the results of other studies~\cite{kuhn2010sp,kuhn2004software,kuhn2002investigation} reporting the degree of interaction occurring in real faults occurring in industrial systems. While not conclusive, the results of this study are interesting because they suggest that the degree of interaction involved in faults might not be as low as previously thought. As a direct result, testing all $4-wise$ combinations might not provide reasonable assurance in terms of fault detection. There is a need to consider ways of studying the use of higher-strength algorithms and tailoring these to the programs considered in this study, which are used in real-time software systems to provide control capabilities in trains. The behavior of such a program depends not only on the choice of parameters but also on providing the right choice of continuous values. By considering the state of the system of the timing information, combinatorial tests might be more effective at detecting faults. Bergstr\"{o}m \textit{et al.}~\cite{bergstrom2017using} indicated that the use of timing information in combinatorial testing for base-choice criterion results in higher code coverage and fault detection. This needs to be further studied by considering the extent to which $t-wise$ can be used in combination with the real-time behavior of the input parameters.

\section{Limitations} \label{limitation}

Our results regarding effectiveness are not based on naturally occurring faults. In our study, we automatically seeded mutants to measure the fault detection capability of the written tests. While it is possible that faults are naturally happening in the industry would yield different results, there are some evidence~\cite{just2014mutants} to support the use of injected faults as substitutes for real faults. Another possible risk of evaluating test suites based on mutation analysis is the {\it equivalent mutant} problem in which these faults cannot show any externally visible deviation. The mutation score in this study was calculated based on the ratio of killed mutants to mutants in total (including equivalent mutants, as we do not know which mutants are equivalent). Unfortunately, this fact introduces a threat to the validity of this measurement. In addition, the results are based on a case study in one company using 37 PLC programs. Even if this number can be considered quite small, we argue that having access to real industrial programs created by engineers working in the safety-critical domain can be representative. More studies are needed to generalize these results to other systems and domains.

Finally, our general clustering problem has also dealt with small scale problems (the largest class diagram is only 31 classes). As the classes get larger, enumeration of the possible solution grows in a factorial manner. With such growth, there could be a potential clustering mismatch. In this case, maximizing MQ can be seen as two conflicting sides of the same coin. On one side of the coin, there is a need to get the largest MQ for better modularization.  On the other side of the coin, automatically maximizing MQ for a large set of classes may be counter-productive (in terms of disrupting the overall architectural package structure of the classes).  In fact, some individual clusters may not be intuitive to programmers at all. For these reasons, there is a need to balance between getting the good enough MQ (i.e., which may not be the best one) and simultaneously obtaining a meaningful set of clusters.

\section{Conclusions} \label{ConclusionSection}
We present Q-EMCQ, a Q-learning based hyper-heuristic exponential Monte Carlo with a counter strategy for combinatorial interaction test generation and show the evaluation results obtained from a case study performed at Bombardier Transportation, a large-scale company focusing on developing industrial control software. The 37 programs considered in this study have been in development and are used in different train products all over the world. The evaluation shows that the Q-EMCQ test generation method is efficient in terms of generation time and test suite size. Our results suggest that combinatorial interaction test generation can achieve high branch coverage. However, these generated test suites do not show high levels of fault detection in terms of mutation score and are more costly (i.e., in terms of the number of created test cases) than manual test suites created by experienced industrial engineers. The obtained results are useful for both practitioners, tool developers, and researchers. Finally, to complement our current work, we have also demonstrated the generality of Q-EMCQ via addressing the maximization problem (i.e., involving the clustering of class diagrams). For future work, we can focus on exploring the adoption of Q-EMCQ for large embedded software both for t-wise test generation as well as its modularization.

\section*{Acknowledgement}

The work reported in this paper is funded by Fundamental Research Grant from the Ministry of Higher Education Malaysia titled: An Artificial Neural Network-Sine Cosine Algorithm) based Hybrid Prediction Model for the production of Cellulose Nanocrystals from Oil Palm Empty Fruit Bunch (RDU1918014). Wasif Afzal is supported by The Knowledge Foundation through 20160139 (TestMine) \& 20130085 (TOCSYC). Eduard Enoiu is funded from the Electronic Component Systems for Eu- European Leadership Joint Undertaking under grant agreement No. 737494 and The Swedish Innovation Agency, Vinnova (MegaM@Rt2).

\bibliographystyle{model1-num-names}
\bibliography{sample.bib}
\end{document}